\documentclass[5p,twocolumn,preprint,times]{elsarticle}
\usepackage{natbib}
\usepackage{graphicx}
\usepackage{epstopdf, epsfig}
\usepackage{amsmath}
\usepackage{dutchcal}
\usepackage{epstopdf}
\usepackage{tikz}
\usepackage{tabularx}
\usepackage{babel}[english]
\usetikzlibrary{shapes,arrows}
\usepackage{array}
\usepackage{verbatim}
\usepackage{comment}
\usetikzlibrary{shapes.geometric}
\newcolumntype{L}[1]{>{\raggedright\let\newline\\\arraybackslash\hspace{0pt}}m{#1}}
\newcolumntype{C}[1]{>{\centering\let\newline\\\arraybackslash\hspace{0pt}}m{#1}}
\newcolumntype{R}[1]{>{\raggedleft\let\newline\\\arraybackslash\hspace{0pt}}m{#1}}
\usepackage{enumerate}
\usepackage{dcolumn}
\usepackage{bm}
\usepackage{xcolor}

\usepackage{multicol}

\usepackage[colorlinks=true,urlcolor=blue]{hyperref} 
\usepackage{xspace}
\usepackage{booktabs}
\usepackage{multirow}
\usepackage{adjustbox}
\usepackage{titlesec}
\usepackage{relsize}
\usepackage{colortbl}
\usepackage{steinmetz}
\usepackage{scalerel}
\usepackage{amssymb}
\usepackage{float}
\usepackage{subcaption}
\usepackage{caption}
\usepackage{cancel}

\usepackage{siunitx}
\usepackage{hyperref}

\journal{International Journal of Heat and Mass Transfer}

\newcommand*{\Nusselt}{\mathrm{Nu}}

\newcommand*{\Rayleigh}{\mathrm{Ra}}
\newcommand*{\Prandtl}{\mathrm{Pr}}

\newcommand*{\Jakob}{\mathrm{Ja}}

\usepackage{nomencl} 

\makenomenclature
\renewcommand*\nompreamble{\begin{multicols}{2}}
	\renewcommand*\nompostamble{\end{multicols}}
\usepackage{framed} 
\usepackage{etoolbox} 
\renewcommand\nomgroup[1]{%
	\item[\bfseries
	\ifstrequal{#1}{A}{Letter symbols}{%
		\ifstrequal{#1}{B}{Greek symbols}{%
			\ifstrequal{#1}{C}{Subscripts and superscripts}{}}}%
	]}

\usepackage{etoolbox}

\renewcommand\nomgroup[1]{%
	\item[\bfseries
	\ifstrequal{#1}{L}{Letter symbols}{%
		\ifstrequal{#1}{G}{Greek symbols}{%
			\ifstrequal{#1}{N}{Non-dimensional numbers}{%
				\ifstrequal{#1}{A}{Abbreviations}{%
					\ifstrequal{#1}{S}{Subscripts and superscripts}{}}}}}%
	]}

\begin{document}

\begin{frontmatter}


    \title{On the complex interplay of temperature, phase change and natural convection in self-pressurization-- an investigation using segregated modeling}


    
    \author[1]{D. Barreiro-Villaverde\corref{cor1}}
    \ead{david.barreiro1@udc.es, david.barreiro@vki.ac.be}
    \author[1]{A. Cantiani}
    \author[1,2,3]{M. A. Mendez}
    \cortext[cor1]{Corresponding author. Currently at Universidade da Coruña, Campus Industrial de Ferrol, CITENI, 15403 Ferrol, Spain.}
    \address[1]{von Karman Institute for Fluid Dynamics, Waterloosesteenweg 72, Sint-Genesius-Rode, Belgium}
    \address[2]{Aerospace Engineering Research Group, Universidad Carlos III de Madrid, Av. de la Universidad 30, 28911 Leganés, Spain}
    \address[3]{Aero-Thermo-Mechanics Laboratory, École Polytechnique de Bruxelles, Université Libre de Bruxelles, Av. Franklin Roosevelt 50, Brussels, 1050, Belgium}
      \date{\today}

    \begin{abstract}
Accurate prediction of self-pressurization in cryogenic tanks requires resolving the coupled effects of heat ingress, natural convection, and phase change. This work introduces a segregated numerical framework in which the liquid and vapor phases are treated with incompressible and compressible solvers, respectively, and the liquid–vapor interface is modeled as a sharp boundary subject to energy-jump conditions derived from first principles, without accommodation or tuning coefficients. Conjugate heat transfer through the tank walls is accounted for by solving the heat-conduction equation in the solid domain rather than prescribing external heat-flux conditions.
The framework is validated against laboratory-scale LN$_2$ and large-scale LH$_2$ experiments, reproducing the spatio-temporal evolution of pressure and temperature without adjustable parameters. In both settings, the simulations identify two distinct regimes in self-pressurization: an initial heating-driven phase that establishes a self-similar temperature profile in the vapor, followed by an evaporation-driven phase in which the pressure rise is governed by the saturation relation. The comparison between these largely different scales motivated a revised scaling for self-pressurization, based on ullage thermodynamics. Finally, the influence of buoyancy was examined by reducing the strength of the gravitational body force, which revealed that natural convection modifies the duration of the transient heating phase but has a limited impact on the long-term pressurization rate. This analysis also clarifies the mechanism controlling the development of thermal stratification in the liquid.
Overall, the segregated approach provides a predictive, parameter-free tool for analyzing cryogenic storage and offers a physically grounded basis for scaling self-pressurization across fluids, geometries, and heat-flux conditions.
    \end{abstract}

\end{frontmatter}

\begin{table*}[htb!]  
    \begin{framed}
        \printnomenclature 
        
        \nomenclature[A]{LH$_2$}{Liquid Hydrogen}
        \nomenclature[A]{LN$_2$}{Liquid Nitrogen}
        \nomenclature[A]{TSR}{Thermal Stratified Region}
        \nomenclature[A]{CFD}{Computational Fluid Dynamics}
        \nomenclature[A]{CHT}{Conjugate Heat Transfer}
        \nomenclature[A]{VOF}{Volume of Fluid}
        \nomenclature[A]{KTG}{Kinetic Theory of Gases}

        \nomenclature[L]{$A$}{Area, \unit{\square\metre}}
        \nomenclature[L]{$c_p$}{Isobaric specific heat, \unit{\joule\per\kilogram\per\kelvin}}
        \nomenclature[L]{$c_v$}{Isochoric specific heat, \unit{\joule\per\kilogram\per\kelvin}}
        \nomenclature[L]{$g$}{Gravitational acceleration, \unit{\metre\per\square\second}}
        \nomenclature[L]{$H$}{Liquid level, \unit{\metre}}
        \nomenclature[L]{$\mathcal{h}$}{Mass-specific enthalpy, \unit{\joule\per\kilogram}} 
        \nomenclature[L]{$\mathcal{L}_v$}{Latent heat of vaporization, \unit{\joule\per\kilogram}} 
        \nomenclature[L]{$m$}{Mass, \unit{\kilogram}}
        \nomenclature[L]{$\dot{m}$}{Mass flow rate / Evaporation rate, \unit{\kilogram\per\second}} 
        \nomenclature[L]{$p$}{Pressure, \unit{\pascal}}
        \nomenclature[L]{$\dot{Q}$}{Heat transfer rate, \unit{\watt}}
        \nomenclature[L]{$\dot{q}$}{Heat flux, \unit{\watt\per\square\metre}}
        \nomenclature[L]{$R$}{Gas constant, \unit{\joule\per\kilogram\per\kelvin}}
        \nomenclature[L]{$T$}{Temperature, \unit{\kelvin}}
        \nomenclature[L]{$t$}{Time, \unit{\second}}
        \nomenclature[L]{$\mathbf{u}$}{Velocity vector, \unit{\metre\per\second}} 
        \nomenclature[L]{$V$}{Volume, \unit{\cubic\metre}}
        \nomenclature[L]{$x, y, z$}{Cartesian coordinates, \unit{\metre}}

        \nomenclature[G]{$\alpha$}{Thermal diffusivity, \unit{\square\metre\per\second}}
        \nomenclature[G]{$\beta$}{Volumetric thermal expansion coefficient, \unit{\per\kelvin}}
        \nomenclature[G]{$\delta_{T}$}{Thermal boundary layer thickness, \unit{\metre}}
        \nomenclature[G]{$\delta_{TSR}$}{Thickness of the thermal stratified region, \unit{\metre}} 
        \nomenclature[G]{$\Delta_{TSR}$}{Temperature difference of the thermal stratified region, \unit{\kelvin}} 
        \nomenclature[G]{$\hat{\Theta}$}{Dimensionless temperature}
        \nomenclature[G]{$\eta$}{Pressurization ratio (Phase change vs. Thermal exp.)} 
        \nomenclature[G]{$\kappa$}{Thermal conductivity, \unit{\watt\per\metre\per\kelvin}}
        \nomenclature[G]{$\mu$}{Dynamic viscosity, \unit{\pascal\second}}
        \nomenclature[G]{$\nu$}{Kinematic viscosity, \unit{\square\metre\per\second}}
        \nomenclature[G]{$\Pi$}{Dimensionless group} 
        \nomenclature[G]{$\rho$}{Density, \unit{\kilogram\per\cubic\metre}}
        \nomenclature[G]{$\sigma$}{Surface tension, \unit{\newton\per\metre}}
        
        \nomenclature[S]{$l$}{Liquid}
        \nomenclature[S]{$v$}{Vapor}
        \nomenclature[S]{$i$}{Interface}
        \nomenclature[S]{$w$}{Wall}
        \nomenclature[S]{$sat$}{Saturation}
        \nomenclature[S]{$s$}{Solid}
        \nomenclature[S]{$0$}{Initial condition}

        \nomenclature[N]{Ra}{Rayleigh number}
        \nomenclature[N]{Pr}{Prandtl number}
        \nomenclature[N]{Nu}{Nusselt number}
        \nomenclature[N]{Ja}{Jakob number}
        
    \end{framed}
\end{table*}

\section{Introduction}\label{sec:introduction}

Accurate prediction of heat and mass transfer is essential for the design of cryogenic storage systems used for liquid hydrogen (LH$_2$), liquefied natural gas (LNG), and related cryogenic fluids. Such systems are increasingly relevant to the energy transition \cite{Squadrito2023} and to applications in the maritime \cite{Fu2023}, aviation \cite{Bicer2017}, and space sectors \cite{Muratov2011,Simonini2024}. The large temperature difference between the stored fluid and its environment drives unavoidable heat ingress through the tank walls, leading to a progressive rise in pressure, commonly referred to as self-pressurization. Mitigation by venting boil-off vapor results in mass and energy losses \cite{Morales2023} and is particularly undesirable in space applications, where venting may induce liquid expulsion \cite{Kartuzova2014,hansen2020}. Understanding the mechanisms governing self-pressurization is therefore essential for the design and operation of long-duration cryogenic storage systems.

Self-pressurization arises from the interaction of several tightly coupled processes, including natural convection, thermal stratification, phase change, and interfacial heat and mass transfer \cite{Wang2024_review}. In the liquid phase, heat ingress through the tank walls generates buoyancy-driven boundary layers that form a thermally stratified region beneath the interface, while maintaining a subcooled bulk below \cite{Sherif1997,Kang2018}. In the vapor phase, temperature and pressure increase more rapidly due to the much lower heat capacity, and interfacial heat transfer can induce evaporation or condensation, directly influencing pressure evolution. Additional complexities stem from recondensation on cold surfaces and from the strong temperature dependence of thermophysical properties under cryogenic conditions.

Despite extensive experimental and numerical studies addressing the influence of tank geometry, heat ingress, and gravity, the mechanisms governing self-pressurization remain incompletely characterized, particularly under non-venting conditions \cite{Wang2024_review}. Experimental investigations are intrinsically challenging due to cryogenic operating conditions and the strong coupling between thermal, hydrodynamic, and phase-change processes. Measurements are typically restricted to global quantities or sparse pointwise data, which limits the ability to isolate individual mechanisms and to infer quantities such as interfacial mass transfer or evaporation rates across different scales without resorting to modeling (e.g. \cite{marques_real_time_2024}). Numerical modeling offers greater diagnostic capability but remains constrained by the treatment of phase change, which is commonly introduced through empirically calibrated interfacial mass-transfer models rather than derived from first principles \cite{Kharangate2017,Zuo2023}. As a result, predictive simulations of self-pressurization continue to rely on case-specific tuning, limiting their applicability beyond validated conditions.

A large body of experimental work has investigated how the pressurization rate depends on parameters such as liquid fill level and heat ingress distribution, yet contrasting trends have been reported. Liebenberg and Edeskuty \cite{Liebenberg1965} observed that increasing the liquid fill level reduced the pressurization rate in a large spherical tank (\(208\,\mathrm{m}^3\)) under spatially uniform heating conditions. In contrast, experiments by Aydelott et al. \cite{aydelott_normal_1967} on a much smaller tank (\(6.37 \times 10^{-3}\,\mathrm{m}^3\)) reported higher pressurization rates at higher fill levels. Besides the possible size effects, the main difference in these two works is in the heating configuration: in \cite{aydelott_normal_1967} the average heat flux imposed on the liquid was kept constant, but the heat input to the vapor region varied with fill level, such that changes in liquid filling were accompanied by changes in the total heat ingress. The same study further showed that heat applied in the vapor region leads to significantly higher pressurization rates than uniform or liquid-only heating, highlighting the importance of the spatial distribution of heat ingress. Consistently, Anderson and Kolar \cite{Anderson1963}, and Tatom et al.\ \cite{Tatom1964} demonstrated that confining heat input to the tank bottom can suppress the development of liquid thermal stratification.

Later work by Aydelott et al.\ \cite{aydelott_effect_1969} showed that the pressurization rate correlates well with the \emph{volumetric} heat flux when identical tanks of different sizes are considered. Hasan \cite{hasan_self-pressurization_1991} and Van Dresar \cite{van_dresar_self-pressurization_1992} investigated the effects of liquid filling, heat flux, and initial conditions in a \(4.9\,\mathrm{m}^3\) ellipsoidal liquid-hydrogen tank. Their experiments documented the spatio-temporal evolution of temperature in both the liquid and vapor phases and have since served as reference data for the validation of numerous theoretical and numerical models. In these studies, the dependence of the pressurization rate on liquid fill level was not monotonic, with intermediate fillings exhibiting the lowest pressurization rates.

More recent investigations have focused on highly instrumented small-scale tanks, enabling detailed characterization of the thermal stratified layer and its evolution under varying fill levels and heat-leak conditions \cite{Seo2010,Kang2018,Perez2021}. These experiments confirmed that thermal stratification plays a central role in self-pressurization and is a primary source of discrepancy between measurements and simplified analytical models, underscoring the need for a deeper understanding of the mechanisms governing its development.

The variability observed across experimental studies reflects the strong dependence of self-pressurization on tank geometry, liquid filling, and the spatial distribution of heat ingress. In this context, theoretical models have been developed to interpret pressurization dynamics using global mass and energy balances. These approaches typically represent interfacial heat and mass transfer through empirical correlations and treat the tank as a lumped system \cite{Wang2024_review,Wang2024_review_2}, thereby providing limited insight into inherently distributed phenomena such as thermal stratification, buoyancy-driven flows, and spatially varying heat transfer within the tank.

Numerical simulations have therefore been increasingly employed to investigate self-pressurization, most commonly using two-phase Volume of Fluid (VOF) formulations \cite{Wang2024_review}, alongside alternative interface-capturing or -tracking approaches such as geometric reconstruction \cite{Malan2021}, level-set methods \cite{Xiao2024}, and Arbitrary Lagrangian--Eulerian techniques \cite{Sim2011}. While these methods differ in how the liquid--vapor interface is represented, they share a common difficulty in the modeling of interfacial heat and mass transfer. In practice, phase change is introduced either through models derived from the Kinetic Theory of Gases (KTG) or through energy-jump conditions imposed at the interface. KTG-based formulations trace their origin to the Hertz--Knudsen relation \cite{Knudsen1967}, with subsequent refinements by Schrage \cite{Schrage1953} and later simplifications by Lee and Tanasawa \cite{Lee1980,Tanasawa1991}. These models rely on accommodation coefficients representing the fraction of molecules that undergo phase change upon impingement at the interface. Despite their physical motivation, the values of these coefficients are typically adjusted over several orders of magnitude \cite{Vaartstra2022}. For example, Kassemi and Kartuzova employed values of \(10^{-3}\,\mathrm{s}^{-1}\) for both evaporation and condensation, whereas Lv et al.\ \cite{Lv2021} adopted coefficients of \(10^{-5}\) for evaporation and \(10^{4}\) for condensation under otherwise identical conditions.

Energy-jump formulations instead compute the phase-change rate directly from interfacial heat fluxes. Although this approach follows directly from energy conservation, its practical implementation within VOF-based frameworks remains challenging due to inaccuracies in interfacial temperature gradients and effective thermal properties near the smeared interface \cite{Zuo2021_CHT}.

These limitations have motivated the development of sharper interface‑capturing methods and segregated solvers. For instance, Malan et al. \cite{Malan2021} applied a geometric‑reconstruction scheme that represents an infinitely thin interface on which energy‑jump conditions are imposed, showing excellent agreement for bubble and droplet dynamics. In the context of cryogenic storage, Sim et al. \cite{Sim2011} modeled two‑phase flow with an Eulerian–Lagrangian formulation, and Xiao et al. \cite{Xiao2024} used a hybrid VOF–level‑set method, although the latter still relies on KTG‑based evaporation models.

In the segregated formulations, the liquid and vapor domains are solved separately with phase‑appropriate equations and coupled through a sharp interface. These approaches improve the resolution of interfacial heat fluxes and reduce numerical stiffness by allowing tailored treatments of compressibility, buoyancy, and turbulence in each phase. attick et al. \cite{Mattick2012} implemented a segregated framework in which phase change is modeled via the Schrage relation, showing satisfactory agreement with large‑scale experiments \cite{Kassemi2023}, yet empirical tuning of interfacial parameters still persists. Recently, Huerta and Vesovic \cite{Huerta2021,Huerta2024} coupled a multidimensional solver for one phase with a reduced‑order model for the other, thereby characterizing buoyancy‑driven flows in isobaric and non‑isobaric processes.

Although segregated formulations have the potential to significantly improve physical consistency of cryogenic tank simulations, the interfacial phase-change rate is still commonly evaluated through kinetic relations or auxiliary assumptions that introduce adjustable parameters or near-equilibrium constraints, rather than being determined directly from the resolved interfacial energy balance.

A further modeling challenge concerns the treatment of heat transfer through the tank walls. Many numerical studies prescribe a uniform heat flux at the internal wall, rather than resolving the conjugate heat transfer (CHT) problem that couples wall conduction with the surrounding fluid domains \cite{Wang2024_review}. This assumption neglects the dependence of the local heat ingress on wall and fluid temperatures and has been shown to overpredict vapor temperatures and pressurization rates, thereby modifying the interfacial heat and mass transfer \cite{Zuo2021_noCHT}. As a result, several studies account for wall heat transfer through simplified boundary-condition models \cite{Liu2019,Huerta2024}, while others solve the wall conduction problem but still impose uniform external heat fluxes \cite{Stewart2016,kassemi_validation_2018,Zuo2021_CHT}. Fully resolving CHT is essential because the spatial distribution of heat ingress strongly influences buoyancy-driven flows, thermal stratification, and ultimately the pressure evolution.

Taken together with the empirical treatment of interfacial phase change, these modeling choices limit the ability of current numerical approaches to predict self-pressurization across different operating conditions. This is reflected in the disparate trends reported in the literature. For example, Lv et al.~\cite{Lv2021} observed lower pressurization rates at higher gravity levels in large tanks, despite the expected enhancement of natural convection, whereas Fu et al.~\cite{Fu2015} reported a negligible influence of gravity in small-scale configurations. Similarly, wall rib structures were found to reduce pressurization rates in short-duration simulations, although their effect on long-term storage remains unclear \cite{Fu2014}. These observations underline the need for modeling frameworks that resolve the coupled mechanisms of heat transfer, buoyancy, and phase change without reliance on ad hoc adjustments.

This paper contributes to the numerical investigation of self-pressurization in cryogenic tanks in two complementary ways. First, a segregated numerical framework is introduced that couples an incompressible buoyant solver for the liquid with a compressible solver for the vapor, and models phase change through energy-jump conditions enforced at a sharp liquid--vapor interface. In contrast to most existing approaches, conjugate heat transfer through the tank walls is fully resolved, allowing the spatial distribution of heat ingress to emerge from the coupled wall--fluid interaction rather than being prescribed. The framework is validated against both laboratory-scale and full-scale experiments using liquid nitrogen and liquid hydrogen, and accurately reproduces the spatio-temporal evolution of pressure and temperature without adjustable parameters.

Second, the developed framework was used to disentangle the respective roles of heat ingress, natural convection, thermal stratification, and interfacial phase change in driving the pressure rise. By systematically varying gravity and thermal boundary conditions, the analysis clarifies the mechanisms linking liquid stratification and evaporation, reconciling competing interpretations reported in the literature—namely, whether evaporation is driven by heat accumulation beneath the interface \cite{Sherif1997} or by the inability of the bulk liquid to track the increasing saturation temperature associated with pressure build-up \cite{Bourgarel1967}. This mechanistic understanding further motivates an alternative scaling of self-pressurization based on the dominant physical regime rather than global heat-input metrics.

The remainder of this paper is organized as follows. Section~\ref{sec:CFD} describes the implementation of our segregated solver, recalling the governing equations in both domains and the implementation of the phase-change formulation, Section~\ref{sec:test_cases} describes the test cases for validation with the scaling laws for comparison, Section~\ref{sec:assessment} presents the assessment and validation of the segregated solver, and Section~\ref{sec:results} investigates the physics of self-pressurization, with special attention on the relative roles of temperature and phase change in the pressurization rate and the effect of natural convection on the system. Finally, Section~\ref{sec:conclusions} concludes recalling the main findings and the future work.

\section{Numerical methodology}\label{sec:CFD}

This section describes the numerical framework used in the present study. The approach is based on a segregated treatment of the liquid and vapor phases, coupled with a conjugate heat transfer formulation for the tank walls. Section~\ref{sec:CFD_model} details the governing equations and numerical treatment of the solid, liquid, and vapor domains. The modeling of phase change at the liquid--vapor interface and the coupling strategy between the two fluid solvers are described in Section~\ref{sec:CFD_coupling}.

\begin{figure}[!htb]
	\centering
	\includegraphics[width=0.95\linewidth]{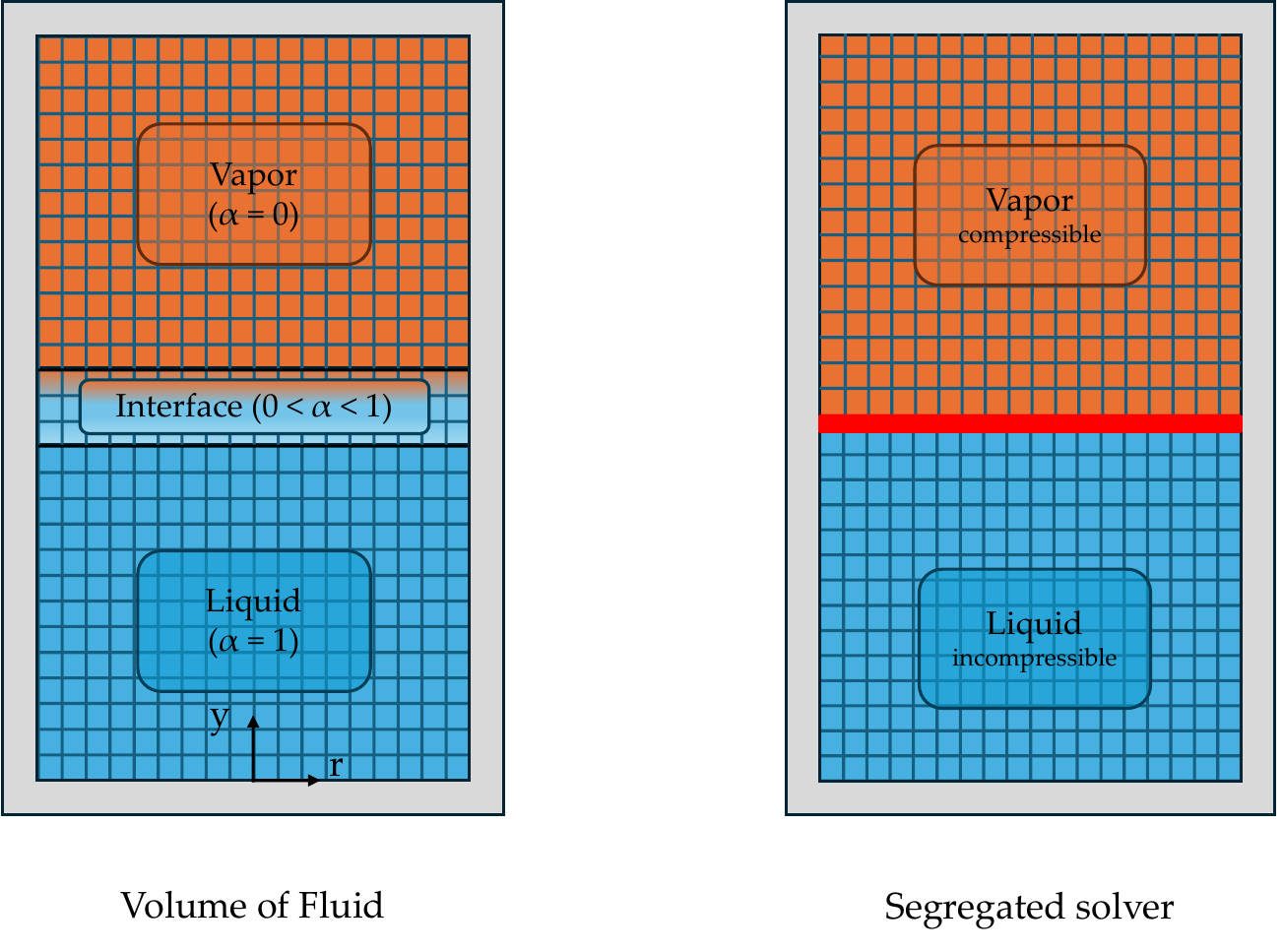}
	\caption{Schematic of the difference between the traditional Volume of Fluid (VOF) approach, on the left and the segregated solver, on the right. The red solid line at the interface in the segregated solver highlights the discrete boundary surface at the liquid–vapor interface where mass and energy exchange due to evaporation and/or condensation is enforced (see Section~\ref{sec:CFD_coupling}).}
	\label{fig:sketch_evaporationModel}
\end{figure}

\subsection{Numerical modeling of the liquid, vapor, and solid}\label{sec:CFD_model}

Figure~\ref{fig:sketch_evaporationModel} illustrates the segregated strategy adopted in this work and contrasts it with conventional monolithic Volume of Fluid (VOF) formulations. In the segregated approach, the liquid--vapor interface is represented as a sharp boundary at which heat and mass transfer are enforced through boundary conditions, ensuring strict conservation and allowing discontinuities in thermodynamic properties to be preserved. Monolithic VOF-based methods, by contrast, represent the interface as a finite-thickness region and introduce phase change through volumetric source terms, which may smear property jumps and require accurate resolution of interfacial gradients.

By solving phase-specific governing equations in separate domains, the sharp-interface formulation enables direct evaluation of interfacial heat fluxes and reduces numerical stiffness associated with diffuse-interface treatments. Although the approach requires communication between the liquid and vapor solvers, this overhead is largely offset by reduced sensitivity to mesh resolution near the interface (see Section~\ref{sec:assessment_comparison}).

For the liquid domain, we employ the \textit{buoyant\allowbreak Boussinesq\allowbreak Pimple\allowbreak Foam} solver from OpenFOAM v2406, which implements the Boussinesq approximation to solve incompressible mass, momentum, and energy equations:

\begin{subequations}
\begin{equation} \label{eq:continuity_liquid}
\nabla \cdot \mathbf{u} = 0,
\end{equation}

\begin{equation}
\begin{split}
\frac{\partial \mathbf{u}}{\partial t} + (\mathbf{u} \cdot \nabla) \mathbf{u} =
    \nabla p_{\text{rgh}} +
    \nabla \cdot \left[ \nu \left( \nabla \mathbf{u} + \nabla \mathbf{u}^{\mathsf{T}} \right) - \frac{2}{3} \nu (\nabla \cdot \mathbf{u}) \mathbf{I} \right] \\
    + \rho \beta (T - T_{\text{ref}})\mathbf{g},
\end{split}
\label{eq:momentum_liquid}
\end{equation}

\begin{equation} \label{eq:energy_liquid}
\frac{\partial T}{\partial t} + \mathbf{u} \cdot \nabla T =
\nabla \cdot \left( \alpha \nabla T \right) \,,
\end{equation}
\end{subequations} where $\mathbf{u}$ is velocity, $\nu$ kinematic viscosity, $\mathbf{I}$ the identity tensor, $T$ temperature, $k$ thermal conductivity, $\rho$ density, $c_p$ specific heat, $\alpha = k / (\rho c_p)$ thermal diffusivity. The buoyancy is modeled via Boussinesq approximation (last term in Equation~\ref{eq:momentum_liquid}), where $\beta$ the thermal expansion coefficient and $T_{\text{ref}}$ the reference temperature. This approximation is valid when $\beta (T - T_{\text{ref}}) \ll 1$, as typically occurs in cryogenic liquids \cite{Ferziger2002}. The pressure term $p_{\text{rgh}} = p - \rho \mathbf{g} \cdot \mathbf{x}$ accounts for hydrostatic effects, with $\mathbf{g}$ gravity and $\mathbf{x}$ position.

The vapor phase is solved using \textit{rhoPimpleFoam}, a transient compressible solver with energy conservation:

\begin{subequations}
\centering
\begin{equation}  \label{eq:continuity_vapor}
\frac{\partial \rho}{\partial t} + \nabla \cdot (\rho \mathbf{u}) = 0 \,,
\end{equation}

\begin{equation}  \label{eq:momentum_vapor}
\begin{split}
\frac{\partial (\rho \mathbf{u})}{\partial t} + \nabla \cdot (\rho \mathbf{u} \mathbf{u}) =
-\nabla p  \\ 
+  \nabla \cdot \left[ \mu \left( \nabla \mathbf{u} + \nabla \mathbf{u}^{\mathsf{T}} \right) - \frac{2}{3} \mu (\nabla \cdot \mathbf{u}) \mathbf{I} \right] + \rho \mathbf{g} \,,
\end{split}
\end{equation}

\begin{equation} \label{eq:energy_vapor}
\begin{split}
\frac{\partial}{\partial t} \left( \frac{1}{2} \rho |\mathbf{u}|^2 + \rho \mathcal{h} \right)
+ \nabla \cdot \left( \frac{1}{2} \rho |\mathbf{u}|^2 \mathbf{u} + \rho \mathcal{h} \mathbf{u} \right) \\
= \frac{dP}{dt} + \nabla \cdot (k \nabla T) + \rho (\mathbf{u} \cdot \mathbf{g}) \,,
\end{split} 
\end{equation}
\end{subequations} where $\mathcal{h}$ is the mass-specific enthalpy and all the other variables are the same as in the liquid solver. 

Although vapor is often modeled as an ideal gas \cite{Wang2024_review}, we here implement the effect of real-gas behavior using the Peng–Robinson (PR) equation of state. The pressure $p$ and temperature $T$, are linked to the gas constant $R$ and the compressibility factor $Z$ as:

\begin{equation}
\rho = \frac{p}{ZRT}, \qquad Z^3 - (1 - B) Z^2 + (A - 2B - 3B^2) Z - (AB - B^2 - B^3) = 0,
\end{equation}with the parameters $A$ and $B$ defined as:

\begin{equation}
\begin{aligned}
A &= \frac{\alpha a p}{R^2 T^2}, \quad  B = \frac{b p}{RT}, \\
a &= 0.45724 \frac{R^2 T_c^2}{p_c}, \quad b = 0.07780 \frac{R T_c}{p_c},
\end{aligned}
\end{equation}

and 

\begin{equation}
\begin{aligned}
\alpha &= \left( 1 + \kappa (1 - \sqrt{T_r}) \right)^2, \\
\kappa &= 0.37464 + 1.54226 \omega - 0.26992 \omega^2,
\end{aligned}
\end{equation} with $T_c$, $p_c$ the critical temperature and pressure, $T_r=T/T_c$ reduced temperature, and $\omega$ the acentric factor. One recovers the ideal gas law if $Z=1$.

Tank walls are treated as no-slip boundaries ($\mathbf{u}=0$). At the liquid--vapor interface, the tangential velocity is set to zero, while the normal velocity component, $u_N$, is prescribed by the evaporation model described in Section~\ref{sec:CFD_coupling}. This assumption is commonly adopted in segregated formulations and is appropriate under the moderate heat-flux conditions considered here, for which interfacial velocities remain small compared to bulk flow velocities \cite{Huerta2024}.

Concerning the modeling of heat ingress, many previous numerical studies assume uniform wall heat fluxes in time and space \cite{Wang2024_review}. This simplification neglects the large disparity in thermal conductivity and heat capacity between liquid and vapor phases, which naturally leads to strongly non-uniform heat-flux distributions along the tank wall. As a consequence, vapor temperatures and pressurization rates are often overestimated when conjugate heat transfer is not resolved \cite{Zuo2021_noCHT}. To mitigate this issue, some authors introduce effective heat-transfer coefficients and prescribe that a fraction of the vapor-side heat flux is absorbed directly at the interface \cite{Huerta2021,Huerta2024}. Others impose time- and space-dependent thermal boundary conditions inferred from experimental measurements \cite{Liu2019}. While these approaches can improve agreement with specific datasets, they rely on empirical assumptions or case-specific inputs and therefore offer limited generality across different geometries, fill levels, and heat-load conditions.

Following established numerical approaches for large-scale problems \cite{Stewart2016,kassemi_validation_2018,Zuo2021_CHT}, we model heat conduction through the tank wall using \textit{solidFoam}, which solves:
\begin{equation}
\rho c_p \frac{\partial T}{\partial t} - \nabla \cdot (k \nabla T) = 0.
\label{eq:heat_solid}
\end{equation}

The external heat flux $\dot{Q}_{ext}$ imposes a boundary gradient:
\begin{equation}
\partial_n T = -\overline{\dot{q}}_{ext}/k, \qquad
\overline{\dot{q}}_{ext} = \dot{Q}_{ext}/A_{ext},
\end{equation}
where $\mathbf{n}$ is the wall-normal vector and $A_{ext}$ the external tank area. The solid–fluid coupling is thus explicit and enforces continuity of temperature and heat flux:
\begin{equation}
k_s \partial_n T|_{s} = k_{l,v} \partial_n T|_{l,v} , \qquad T_s = T_{l,v},
\end{equation}
with $s$, $l$, and $v$ denoting solid, liquid, and vapor sides. An implicit approach was tested but halved computational speed without improving accuracy. The coupling direction assigns temperature from solid to fluid and heat flux from fluid to solid, as this is numerically more stable for low-conductivity fluids \cite{Verstraete2016}. The solid solver is coupled with the fluid solvers in every time step to ensure stability.

The cases are simulated as laminar, following evidence that this approach remains accurate even for modified Rayleigh numbers $Ra^* = g \beta \dot{q} L^4 \rho c_p \nu^{-1} k^{-2}$ up to $\mathcal{O}(14)$ in large-scale tanks \cite{kassemi_validation_2018}. Turbulent models tend to overpredict mixing, leading to unrealistically low vapor temperatures, enhanced vapor-side heat transfer, and consequently an underestimation of ullage pressurization.

All simulations were performed in an axisymmetric configuration using fully hexahedral meshes generated with \textit{blockMesh}. For the laboratory-scale cylindrical tank, a uniform grid spacing was employed throughout the domain. The large-scale ellipsoidal tank was discretized using an O-grid topology with progressive refinement near the tank walls, enabling accurate resolution of thermal and momentum boundary layers while maintaining a reasonable computational cost. The reference wall-normal grid spacing used in the mesh-sensitivity study presented in Section~\ref{sec:assessment_comparison}; this was selected based on established correlations for the thermal boundary layer thickness along a heated vertical wall. Since the thermal boundary layer is more restrictive than the momentum boundary layer for $\Prandtl > 1$, its thickness is estimated as
\begin{equation}
\delta_T = 5 L \, \Rayleigh_L^{*-1/4} \Prandtl^{-1/2},
\end{equation}
where $\Prandtl = \nu / \alpha$ denotes the Prandtl number. For the large-scale configuration, taking $L = 0.96\,\mathrm{m}$ (the distance between the tank top or bottom and the liquid--vapor interface) and an external heat flux $\dot{q}_{\mathrm{ext}} = 3.5\,\mathrm{W/m}^2$, this estimate yields $\delta_T \approx 1.6\,\mathrm{mm}$ in the vapor and $0.33\,\mathrm{mm}$ in the liquid. While local effects such as wall curvature, the presence of the thermal stratified region (TSR), and spatial variations in heat flux may locally alter boundary-layer development, this estimate provides a consistent baseline for resolving wall-normal gradients.

\subsection{Phase change formulation}\label{sec:CFD_coupling}

In the proposed segregated framework, heat and mass transfer due to evaporation are imposed as boundary conditions at the interface, rather than as volumetric source terms in monolithic approaches. This strategy ensures global and local (liquid and vapor) conservation and treats the interface as a discrete surface, enabling direct use of the energy-jump method to compute the evaporation rate $\dot{m}$ from the heat flux across an interfacial face $i$ of area $A_i$:

\begin{equation}
\dot{m}_i = \frac{\dot{Q}_{i}}{\mathcal{L}_v} \quad \,, \text{with } \quad \dot{Q}_{i} = (\dot{q}_{i,v} - \dot{q}_{i,l}) A_i \,,
\end{equation} where $\mathcal{L}_v$ is the latent heat, and $\dot{q}_{i,v} = k_v \partial_n T|_{i,v}$, $\dot{q}_{i,l} = k_l \partial_n T|_{i,l}$ are the heat fluxes from vapor and liquid, respectively, along the normal $\mathbf{n}_i$. The interface is assumed flat and stationary during self-pressurization to reduce computational cost without significantly affecting accuracy \cite{Zuo2023}.

Mass transfer is applied through a normal velocity boundary condition:
\begin{equation} \label{eq:evaporation_u}
u_n = \frac{\dot{m}_i}{\rho_{l,v} A_i} = \frac{\dot{q}_{i,v} - \dot{q}_{i,l}}{\mathcal{L}_v\rho_{l,v}}.
\end{equation}

Similarly, our segregated model enforces thermodynamic equilibrium at the interface, setting the interface temperature to the saturation temperature corresponding to the vapor pressure:
\begin{equation} \label{eq:tsat}
T_{i,v} = T_{i,l} = T_{sat}(p_v) ,.
\end{equation}

In contrast, VOF formulations coupled with kinetic-theory-based phase-change models compute evaporation rates from deviations from equilibrium, such that the interface temperature is not constrained to be uniform along the interface or equal to the saturation temperature.

Coupling between the liquid and vapor solvers was handled using the \textit{preCICE} library \cite{preCICE,preCICE_openFOAM}, while the phase-change calculations were performed by an external Python script. At each coupling step, the script read the vapor pressure and interfacial heat fluxes, updated the interface temperature according to Equation~\ref{eq:tsat}, and imposed the corresponding mass transfer through Dirichlet velocity boundary conditions (Equation~\ref{eq:evaporation_u}). These coupling updates were applied at intervals larger than the solver time step, based on the characteristic time scale of each case, which minimized communication overhead and improved computational efficiency.

Thermophysical properties of the cryogenic fluids were obtained from the open-source \textit{CoolProp} library \cite{CoolProp} and the National Institute of Standards and Technology (NIST) database \cite{NIST}. In OpenFOAM, these properties were represented using temperature-dependent polynomial fits.

\section{Selected test cases and dimensionless analysis}\label{sec:test_cases}

This section describes the experimental configurations reproduced numerically in this study, with emphasis on tank geometry and size, working fluids, and initial and boundary conditions. The laboratory-scale experiments of Perez et al.~\cite{Perez2021} are introduced in Section~\ref{sec:test_cases_Perez}, while the large-scale campaign conducted by Hasan, Van Dresar, and co-workers~\cite{hasan_self-pressurization_1991,van_dresar_pressurization_1993} is presented in Section~\ref{sec:test_cases_Hasan}. Section~\ref{sec:test_cases_scaling} reviews the dimensionless groups governing self-pressurization, with the dual objective of identifying the dominant physical mechanisms and comparing their relative importance across laboratory- and large-scale conditions. On this basis, a revised scaling is proposed that reflects the dominant processes observed in the simulations.

\subsection{Laboratory-scale test case}\label{sec:test_cases_Perez}

The experiments of Perez et al.~\cite{Perez2021} investigated self-pressurization and venting in an upright cylindrical tank of volume $6.73\,\mathrm{L}$ ($200\,\mathrm{mm}$ diameter $\times$ $213\,\mathrm{mm}$ height) filled with liquid nitrogen (LN$_2$). The test cell was constructed from $1.6\,\mathrm{mm}$-thick 316 stainless steel and was enclosed within a copper can housed in a vacuum chamber ($p < 2\,\mathrm{kPa}$), which was itself surrounded by a concentric vacuum-insulated stainless-steel can. External heat input was provided by electrical heaters, and the entire assembly was immersed in a liquid-nitrogen dewar, enabling precise thermal control. The test cell was instrumented with 35 temperature sensors uniformly distributed throughout the domain and a pressure transducer. Data were recorded at 5-minute intervals.

Prior to each experiment, the test cell was filled with LN$_2$ to the prescribed liquid level and allowed to stabilize overnight until thermal equilibrium was reached at approximately $101\,\mathrm{kPa}$ and $77\,\mathrm{K}$. The heaters were then activated, and the experiment proceeded in two stages: an initial self-pressurization phase with the cell sealed, followed by a venting phase in which a pressure-regulating valve maintained a constant pressure while the boil-off rate was measured. In the present study, only the self-pressurization phase was considered, focusing on the case with an $88\,\%$ liquid fill level and a total heat input of $\dot{Q} = 2.2\,\mathrm{W}$.

\subsection{Large-scale test case}\label{sec:test_cases_Hasan}

The experiments conducted by Hasan \cite{hasan_self-pressurization_1991} and Van Dresar \cite{van_dresar_self-pressurization_1992} were designed to reproduce storage conditions representative of spaceflight applications. The test experiments were carried out on a $4.89\,\mathrm{m}^3$ ellipsoidal tank, with major and minor diameters of $2.2\,\mathrm{m}$ and $1.8\,\mathrm{m}$, respectively, filled with liquid hydrogen (LH$_2$). The tank was fabricated from chemically milled 2219 aluminum and insulated using a high-performance multilayer insulation system. It was suspended within a large vacuum chamber and surrounded by a $4.0\,\mathrm{m}$-diameter cryoshroud, which allowed precise control of the net heat flux imposed on the tank.

The tank was tested at liquid fill levels of $83\,\%$, $49\,\%$, and $29\,\%$ by volume and subjected to external heat fluxes $\dot{q}_{\mathrm{ext}}$ of $0.35$, $2.0$, and $3.5\,\mathrm{W/m^2}$. Instrumentation included silicon diode thermometers to measure temperatures in the liquid, vapor, and tank walls, a capacitance probe to determine the liquid level, and pressure transducers to monitor the tank pressure. Experimental data were recorded at 30-minute intervals. Each test was preceded by a boil-off preconditioning phase, during which the tank was filled to approximately $95\,\%$ and vented until steady boil-off rates and stable wall temperatures were achieved. In the present work, the case corresponding to a $49\,\%$ liquid fill level and an imposed heat flux of $\dot{q}_{\mathrm{ext}} = 3.5\,\mathrm{W/m^2}$ from Van Dresar et al.~\cite{van_dresar_self-pressurization_1992} was reproduced. This configuration has also been used as a benchmark in previous numerical studies, including those of Kassemi et al.~\cite{kassemi_validation_2018} and Stewart and Moder~\cite{Stewart2016}, which employed VOF-based solvers with conjugate heat transfer and kinetic-theory-based evaporation models.

\subsection{Scaling considerations} \label{sec:test_cases_scaling}

To interpret self-pressurization across different tank sizes, fluids, and operating conditions, and to assess the relative importance of heat transfer, buoyancy, and phase change, the results are presented in dimensionless form using two alternative scaling approaches. The corresponding reference quantities are summarized in Table~\ref{tab:characteristic_scales} and are used to non-dimensionalize all variables, such as time $\hat{t}=t/[t]$, temperature differences $\widehat{\Delta T}=\Delta T/[\Delta T]$, and evaporation or condensation rates $\widehat{dm_v/dt}=(dm_v/dt)/[dm_v/dt]$.

The first scaling approach follows the formulation proposed by Marques et al.~\cite{Marques2025} (first column of Table~\ref{tab:characteristic_scales}). This scaling is based on the assumption that wall-driven natural convection in the liquid is the primary mechanism governing heat transfer, and that evaporation acts as a secondary process converting the liquid sensible heat into vapor mass and pressure rise. Accordingly, the characteristic length scale is taken as the liquid height $H_l$, and the characteristic time scale $[t]$ is defined by thermal diffusion in the liquid phase.

The reference temperature scale $[\Delta T]$ is obtained from a global energy balance between the imposed heat ingress and the sensible energy stored in the combined liquid--vapor system. The characteristic evaporation rate is then defined as $[dm_v/dt]=[\Delta m]/[t]$, where the characteristic mass variation
$
[\Delta m] = V_{v,0}(\overline{\rho}_{l,0}-\overline{\rho}_{v,0})
$ represents the maximum possible change in ullage mass. Reference rates for pressure and temperature follow from the generic relation $[d\phi/dt]=[\phi]/[t]$ applied to the corresponding base scales.

\begin{table}[t]
\centering
\caption{Characteristic scales for self-pressurization in Marques et al. \cite{Marques2025} and those proposed in this work.}
\label{tab:characteristic_scales}
\renewcommand{\arraystretch}{1.6}
\setlength{\tabcolsep}{5pt}
\begin{tabular}{@{}c | c c@{}}
Quantity & Marques et al. \cite{Marques2025} & Present work \\
\midrule
$[L]$            
  & $H_l$ 
  & $H_v$ \\[0.2em]
$[t]$            
  & $\dfrac{[L]^2}{\bar{\alpha}}$ 
  & $\dfrac{\overline{\rho}_{v,0} c_{p,v} [\Delta T]}{\dot{q}_a A_{v}}$ \\[0.2em]
$[T]$     
  & $T_{sat}(p_0)$ 
  & $T_{sat}(p_0)$ \\[0.2em]
$[\Delta T]$     
  & $\dfrac{\dot{q}_a A_w [t]}{m_0 \bar{C}_p}$ 
  & $\dfrac{\dot{q}_a [L]}{k_v}$ \\[0.2em]
$[dm/dt]$        
  & $\dfrac{(\bar{\rho}_{l,0} - \bar{\rho}_{v,0}) V_{v,0}}{[t]}$ 
  & $\dfrac{\dot{q}_a A_{w,v}}{\mathcal{L}}$ \\[0.2em]
$[dp/dt]$        
  & $\dfrac{(\bar{\rho}_{l,0} - \bar{\rho}_{v,0})}{[t]}\dfrac{\partial p}{\partial \rho}_T$ 
  & $\left.\dfrac{\partial p}{\partial \rho}\right|_{T,0} 
     \dfrac{\dot{q}_a A_{w,v}}{V_{v,0}\,\mathcal{L}}$ \\[0.2em]
 $[dT/dt]$        
  & $\dfrac{[T]}{[t]}$ 
  & $\left.\dfrac{dT_{\text{sat}}}{dp}\right|_0 \,[dp/dt]$ \\
\end{tabular}
\end{table}

In this work, an alternative scaling is proposed (second column of Table~\ref{tab:characteristic_scales}) that is instead based on quantities associated with the ullage volume. This approach is motivated by the long-term behavior observed in the simulations, in which pressure rise is controlled by the conversion of external heat ingress into vapor sensible heating and evaporation, rather than by transient liquid-side convection.

The characteristic time scale $[t]$ is therefore defined from the rate at which the imposed heat flux increases the sensible energy of the vapor. The associated temperature scale $[\Delta T]$ is obtained by assuming that heat transfer within the ullage is dominated by conduction over a characteristic length $H_v$. Under these assumptions, the reference evaporation rate is defined by assuming that, in the long-term limit, the heat input to the ullage is entirely consumed by phase change. The corresponding pressure and temperature rates are then obtained directly from the vapor equation of state and the saturation relation.

\begin{table*}[t]
\centering
\caption{Relevant dimensionless groups in self-pressurization according to Marques et al.\ \cite{Marques2025}. The laboratory \cite{Perez2021} and large-scale \cite{hasan_self-pressurization_1991,van_dresar_self-pressurization_1992} tests are compared using their expressions.}
\label{tab:dimensionless_analysis}
\begin{tabular}{@{}ccccc@{}}
\toprule
Group & Definition & Expression & Lab-scale \cite{Perez2021} & Large-scale \cite{hasan_self-pressurization_1991,van_dresar_self-pressurization_1992} \\
\midrule
$\Pi_3^{l,w}=\Nusselt$  
  & $\dfrac{h_{l,w}[L]}{\kappa_l}$ 
  & $\dfrac{h_{l,w} H_l}{\kappa_{l}}$ 
  & 7.15e2  
  & 6.32e3 \\
$\Pi_4^{l}=\Rayleigh$  
  & $\dfrac{g\beta_f[\Delta T][L]^3}{\nu_l \alpha_l}$ 
  & $\dfrac{g\beta_f \dot{q}_a A_w H_l^5}{\nu_l  \alpha_l m_0 \overline{c}_p \overline{\alpha}}$ 
  & 1.68e11 
  & 1.61e14 \\
$\Pi_5^l=\Prandtl$  
  & $\dfrac{\nu_l}{\alpha_l}$ 
  & $\dfrac{\nu_l}{\alpha_l}$ 
  & 2.27 
  & 1.27 \\
$\Pi_7^{l,w}= \Jakob$  
  & $\dfrac{\kappa_l [\Delta T]}{\mathcal{L}_v [L] [dm_v/dt]}$ 
  & $\dfrac{\kappa_l \dot{q}_a A_w H_l^2}{\mathcal{L}_v m_0\overline{c}_p \overline{\alpha}^2 (\overline{\rho}_{l,0}-\overline{\rho}_{v,0})V_{v,0}}$ 
  & 4.40e1 
  & 1.04 \\
$\Pi_{9}^{v}$  
  & $\dfrac{\dot{q}_a A_{w,v}}{[m_v] c_{p,v} [dT/dt]}$ 
  & $\dfrac{\dot{q}_{a} A_{w,v} H_l^2}{m_{v,0} c_{p,v} T_{sat}(p_0)\overline{\alpha}}$ 
  & 1.45e1 
  & 4.46e1 \\
$\Pi_{9}^{l}$  
  & $\dfrac{\dot{q}_a A_{w,l}}{[m_l] c_{p,l} [dT/dt]}$ 
  & $\dfrac{\dot{q}_{a} A_{w,l} H_l^2}{m_{l,0} c_{p,l} T_{sat}(p_0)\overline{\alpha}}$ 
  & 1.02e-1 
  & 1.01 \\
$\Pi_3^{l}\Pi_7^{l}$ 
  & -- 
  & $\dfrac{h_l \dot{q}_a A_w H_l^3}{\mathcal{L}_v m_{0,l} \overline{c}_p \overline{\alpha}^2  (\overline{\rho}_{l,0}-\overline{\rho}_{v,0})V_{v,0}}$ 
  & 3.14e4 
  & 6.59e3 \\
\bottomrule
\end{tabular}
\end{table*}

To assess the two scaling approaches, the results presented in Section~\ref{sec:results_flow} are shown in both dimensional form (bottom axes) and non-dimensional form (top axes). The liquid-based scaling of Marques et al.~\cite{Marques2025} is used to define dimensionless variables denoted by a hat $\hat{\cdot}$, while the ullage-based scaling proposed here employs the same notation supplemented by a subscript $\hat{\cdot}_*$.

Following the model-based scaling of Marques et al.~\cite{Marques2025}, the dimensionless groups obtained from the non-dimensional governing equations are listed in Table~\ref{tab:dimensionless_analysis}. In this framework, the evaporation and pressurization rates are primarily governed by the Nusselt number $\Pi_3^{l,w}$ and the Jakob number $\Pi_7^{l,w}$, which quantify wall-to-liquid convective heat transfer and the ratio of sensible to latent heat associated with phase change, respectively. For consistency, the enumeration of the dimensionless groups follows that of \cite{Marques2025}. The Nusselt number depends on the Rayleigh number $\Pi_4^l$, which characterizes the intensity of buoyancy-driven convection in the liquid, and on the Prandtl number $\Pi_5^l$, which reflects the relative importance of momentum and thermal diffusion. In contrast, bulk temperature variations are governed by the groups $\Pi_9^{v}$ and $\Pi_9^{l}$, which describe how the imposed heat flux is converted into sensible heating of the vapor and liquid phases, respectively.

Using the reference scales of Marques et al.~\cite{Marques2025}, the laboratory- and large-scale experiments are compared in Table~\ref{tab:dimensionless_analysis}. Both configurations operate in convection-dominated regimes, as indicated by the large Rayleigh and Nusselt numbers. This effect is particularly pronounced in the large-scale case, where the Rayleigh number exceeds that of the laboratory-scale configuration by approximately three orders of magnitude and lies well above the critical value $\Rayleigh_{\mathrm{cr}} \sim \mathcal{O}(10)$, indicating strong buoyancy-driven mixing. The Jakob number is significantly larger in the laboratory-scale experiment, suggesting that a greater fraction of the wall heat input is stored as sensible energy rather than consumed by evaporation. Finally, the values of $\Pi_9^{v}$ and $\Pi_9^{l}$ indicate a much faster temperature rise in the vapor than in the liquid, reflecting the lower thermal capacity of the ullage, particularly in the laboratory-scale configuration.

\section{Assessment of the segregated solver}
\label{sec:assessment}

Before examining the physical mechanisms underlying self-pressurization, the numerical framework is assessed in terms of interface treatment, phase-change modeling, and predictive capability. Section~\ref{sec:assessment_comparison} compares the proposed segregated approach with the commonly used combination of Volume of Fluid and Kinetic Theory of Gases (VOF+KTG). Section~\ref{sec:assessment_validation} evaluates the solver predictions against experimental data for both laboratory-scale and large-scale test cases, assessing its ability to reproduce the observed pressure and temperature evolution across different regimes.

\subsection{Benchmarking against VOF+KTG models: limitations, computational cost and mesh convergence}\label{sec:assessment_comparison}

The large-scale test case is used here to compare the proposed segregated solver with the widely adopted Volume of Fluid formulation coupled with Kinetic Theory of Gases (VOF+KTG). The comparison focuses on differences in interface representation, phase-change modeling, computational cost, and sensitivity to mesh resolution. For this purpose, a VOF+KTG solver based on \textit{compressibleInterFoam} coupled with the Lee phase-change model \cite{Lee1980} is employed. In this formulation, the evaporation rate $\dot{m}$ is given by

\begin{equation} \label{eq:lee_model}
\dot{m} = C \, \rho \, \frac{T - T_{sat}}{T_{sat}} \,,
\end{equation} where $C$ is the accommodation coefficient, which physically governs the rate at which the liquid–vapor interface approaches thermal equilibrium. 

\begin{figure}[h]
\centering
\includegraphics[width=0.95\linewidth]{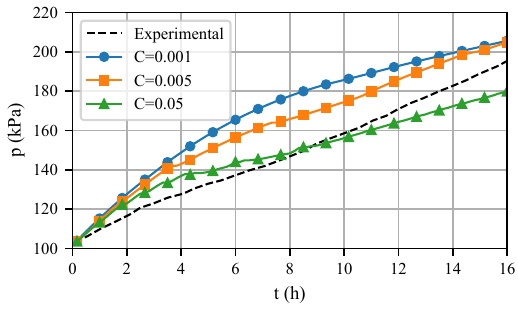}
\caption{Pressurization curve $p(t)$ from large-scale experiments contrasted with numerical simulations using the VOF approach and Lee phase-change model (Equation~\ref{eq:lee_model}) for various accommodation coefficients $C$. The markers are shown only for plotting purposes.}
\label{fig:Hasan_KTGcoeffs}
\end{figure}

As discussed in the introduction, the predictive capability of KTG-based models depends on the choice of the accommodation coefficient, which must typically be calibrated against experimental data. Kassemi and Kartuzova \cite{kassemi_effect_2016} observed that, for long-term storage scenarios, pressurization curves are relatively insensitive to the accommodation coefficient. However, the results displayed in Figure~\ref{fig:Hasan_KTGcoeffs} obtained with our implementation of the VOF+KTG model show differences exceeding $20 , \mathrm{kPa}$ within a narrow range, in line with the deviations reported in the literature \cite{Kharangate2017, Lv2021, Zuo2023, Wang2024_review}.

\begin{figure}[!htb]
\includegraphics[width=0.95\linewidth]{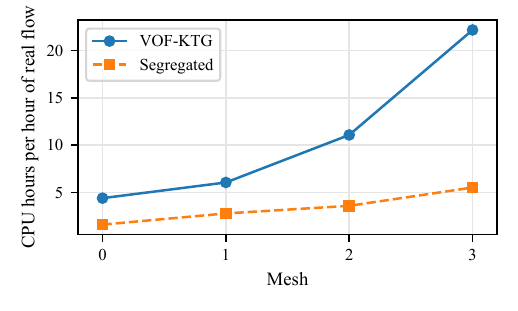}
\caption{Comparison of computational cost of simulating the large-scale test using the segregated solver and VOF+KTG techniques across four mesh resolutions. Mesh sizes: $M_0 = 7.5\, k$, $M_1 = 16.7\, k$, $M_2 = 29.9\, k$, $M_3 = 66.7\, k$ cells.}
\label{fig:Hasan_compCost}
\end{figure}

The segregated solver exhibits improved efficiency and scalability, as demonstrated by the computational cost analysis shown in Figure~\ref{fig:Hasan_compCost} for four mesh resolutions. This advantage stems from tailoring the governing equations to the dominant physics in each phase, with incompressible formulations in the liquid and compressible formulations in the vapor, while avoiding the solution of a volume-fraction transport equation. By solving the liquid and vapor subdomains separately and exchanging information only at the interface, the approach leads to smaller, phase-specific linear systems, in contrast to monolithic formulations that assemble all phases into a single system. The resulting reduction in matrix size outweighs the additional inter-solver communication required to evaluate interfacial heat fluxes, evaporation rates, and saturation temperatures at each coupling step. This makes large-scale configurations and long-duration simulations computationally tractable. Additional performance gains are expected from future implementation of waveform time interpolation \cite{ruth2021}, which would reduce coupling frequency while maintaining stability.

\begin{figure}[!htb]
    \centering
    \begin{subfigure}[b]{0.95\linewidth}
        \includegraphics[width=\linewidth]{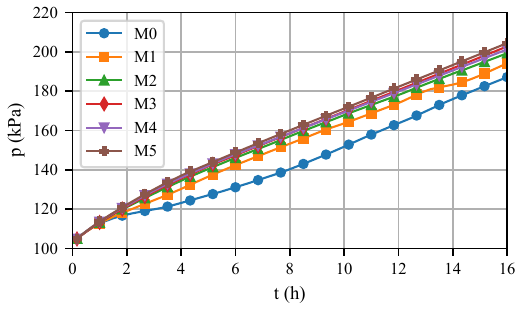}
        \caption{}
        \label{fig:Hasan_meshDependency_p}
    \end{subfigure}
    \hfill
    \begin{subfigure}[b]{0.95\linewidth}
        \includegraphics[width=\linewidth]{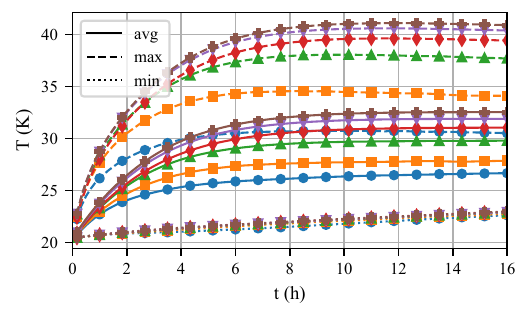}  
        \caption{}
        \label{fig:Hasan_meshDependency_T}
    \end{subfigure}
    \caption{
        Mesh sensitivity analysis for the segregated solver in the large-scale test in terms of pressure (a) and temperature evolution (b) in the vapor phase for various mesh resolutions: $M0 \, (\Delta x = 10\, \mathrm{mm})$ with $7.4$k cells, $M1\, (5\, \mathrm{mm})$ with $17.2$k cells, $M2\, (2\, \mathrm{mm})$ with $39.6$k cells, $M3\, (1\, \mathrm{mm})$ with $61.4$k cells, $M4\, (0.5\, \mathrm{mm})$ with $84.62$k cells, and $M5\, (0.25\, \mathrm{mm})$ with $107.6$k cells. The markers are shown only for plotting purposes.}
    \label{fig:Hasan_meshDependency}
\end{figure}

Mesh convergence of the segregated solver is assessed in Figure~\ref{fig:Hasan_meshDependency}, which compares the pressurization history and temperature evolution obtained with different mesh resolutions. The wall-normal cell size for each mesh is indicated in the figure caption. In all cases, a fixed time step of $1\,\mathrm{ms}$ is used, ensuring that the Courant number remains below $0.5$. Figure~\ref{fig:Hasan_meshDependency_p} shows excellent convergence of the pressurization curves, with meshes $M3$ to $M5$ nearly collapsing onto a single solution. Although coarse meshes ($M0$ to $M2$) exhibit unacceptable offsets at the end of the simulation, they nonetheless predict the long-term pressurization rate reasonably well once the temperature fields reach a quasi-steady state (Figure~\ref{fig:Hasan_meshDependency_T}, $t \gtrsim 8\,\mathrm{h}$).

\begin{figure}[!htb]
\includegraphics[width=0.95\linewidth]{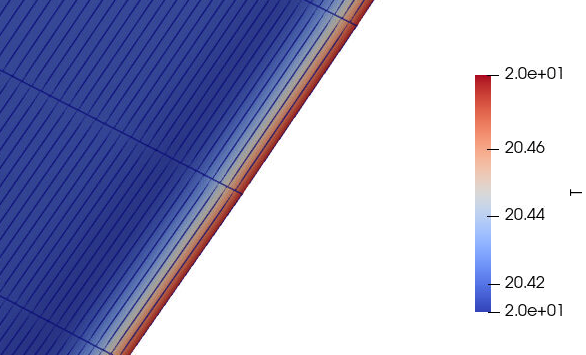}
\caption{Detailed visualization of the mesh in the thermal boundary layer in the large-scale test case.}
\label{fig:Hasan_BL_OF}
\end{figure}

The residual error observed for coarse meshes is attributed to an underestimation of the ullage heat ingress caused by insufficient resolution of the near-wall thermal boundary layers. Small but noticeable differences in temperature evolution persist among meshes $M3$, $M4$, and $M5$, even when the corresponding pressurization curves appear nearly identical. Based on this analysis, mesh $M4$ is selected for the simulations presented in the following sections. This resolution provides approximately five grid points across the liquid-side thermal boundary layer, as illustrated in Figure~\ref{fig:Hasan_BL_OF}. An analogous mesh-convergence study was conducted for the laboratory-scale configuration and is omitted here for brevity.

In summary, this section highlights key limitations of conventional VOF+KTG formulations and demonstrates that the proposed segregated solver enables efficient and mesh-independent simulations of self-pressurization. The approach allows boundary layers on the order of millimeters to be resolved within meter-scale domains, while maintaining millisecond time steps over simulations spanning several hours, making it well suited for large-scale and long-duration cryogenic storage problems.

\subsection{Validation with experiments}\label{sec:assessment_validation}

The segregated solver is validated against the laboratory-scale experiments of Perez et al.~\cite{Perez2021} and the large-scale tests of Hasan, Van Dresar, and co-workers~\cite{hasan_self-pressurization_1991,van_dresar_pressurization_1993}.

For the laboratory-scale configuration, the flow fields were initialized as uniform, except for the ullage temperature, which followed two different initialization strategies. The first, referred to as \emph{Segregated}, assumed fully isothermal conditions in both phases, with $T = T_{\mathrm{sat}}(p_0=101\,\mathrm{kPa})$. The second strategy, denoted \emph{Segregated Prec.}, employed a thermally preconditioned state obtained by initializing the system at $p=75\,\mathrm{kPa}$ and $T=T_{\mathrm{sat}}(75\,\mathrm{kPa})$, thereby introducing an initial thermal stratification representative of the experimental conditions at $t=0$, when the pressure inside the tank is $p_0 = 101 \, kPa$, as it is expected to take place in the experiments.

\begin{figure}[!htb]
	\centering
	\begin{subfigure}[a]{0.99\linewidth}
		\includegraphics[width=\linewidth]{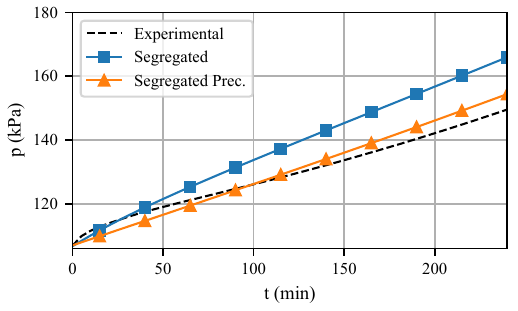}
		\caption{}
		\label{fig:perez_validation_p}
	\end{subfigure}
	\hfill
	\begin{subfigure}[b]{0.99\linewidth}
		\includegraphics[width=\linewidth]{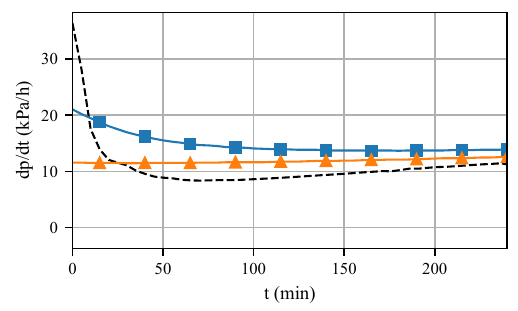}
		\caption{}
		\label{fig:perez_validation_gradP}
	\end{subfigure}
	\caption{Validation of the segregated solver against the laboratory-scale experiments of Perez et al. \cite{Perez2021}: (a) temporal evolution of the tank pressure $p$, and (b) the pressurization rate $dp/dt$. Two initialization strategies are included: an isothermal initialization (Segregated) and a preconditioned state (Segregated Prec.) that accounts for the initial thermal stratification present in the experiments. The markers are shown only for plotting purposes.}
	\label{fig:perez_validation}
\end{figure}

\begin{figure}[!htb]
	\centering
	\begin{subfigure}[a]{0.99\linewidth}
		\includegraphics[width=\linewidth]{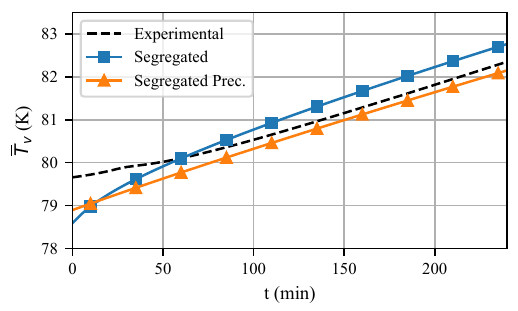}
		\caption{}
		\label{fig:perez_validation_T}
	\end{subfigure}
	\hfill
	\begin{subfigure}[b]{0.99\linewidth}
		\includegraphics[width=\linewidth]{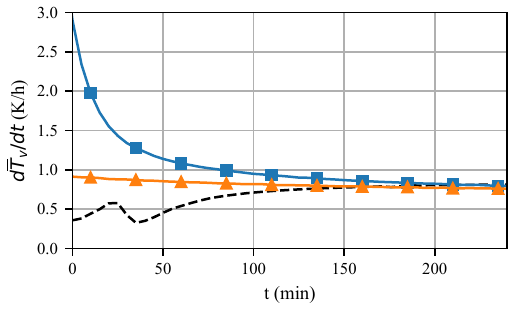}
		\caption{}
		\label{fig:perez_validation_gradT}
	\end{subfigure}
	\caption{Temporal evolution of the mean temperature in the vapor $\overline{T}_v(t)$ (a) and its temporal derivative $d \overline{T}_v /dt$ (b) reported in the experiments of Perez et al. \cite{Perez2021} and the predictions of our segregated solver. The markers are shown only for plotting purposes.}
	\label{fig:perez_validation_temperatures}
\end{figure}

Figure~\ref{fig:perez_validation} presents the temporal evolution of the pressure and pressurization rate, while Figure~\ref{fig:perez_validation_temperatures} shows the corresponding evolution of the mean ullage temperature $\overline{T}_v$ and its rate of change. The mean vapor temperature was computed as a spatial average at the same locations as the experimental probes,
\begin{equation}
  \overline{T}_v (t) = \frac{1}{N_{\mathrm{exp}}} \sum_{i=1}^{N_{\mathrm{exp}}} T(x_i,t),
\end{equation}
where $N_{\mathrm{exp}}$ denotes the number of temperature sensors located in the vapor.

Overall, the agreement between simulations and experiments is satisfactory, particularly during the later stages of the tests when the system becomes insensitive to the initial conditions~\cite{hasan_self-pressurization_1991}. The preconditioned initialization yields improved agreement during the early transient, suggesting that the experimental configuration is more representative of a thermally stratified initial state than a fully isothermal one. The isothermal initialization leads to an overprediction of the early pressurization rate, resulting in an accumulated pressure offset of approximately $20\,\mathrm{kPa}$ at the end of the test. The excellent agreement in both pressure and temperature rates during the quasi-steady phase indicates that the remaining discrepancies primarily originate from uncertainties in the initial thermal state rather than from deficiencies in the model formulation. Additional differences may arise from idealizations in the numerical setup, such as the assumption of uniform external heat flux and constant wall thickness, whereas the experimental apparatus exhibits more complex geometric and thermal features.

\begin{figure}[h]
	\centering
	\begin{subfigure}[a]{0.95\linewidth}
		\includegraphics[width=\linewidth]{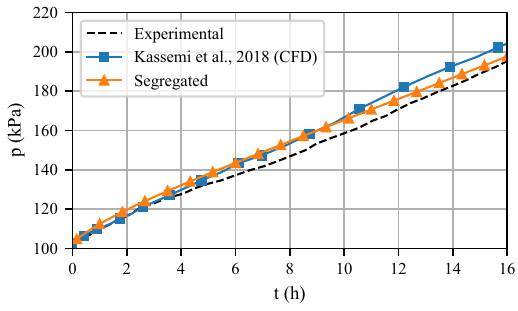}
		\caption{}
		\label{fig:hasan_validation_p}
	\end{subfigure}
	\hfill
	\begin{subfigure}[b]{0.95\linewidth}
		\includegraphics[width=\linewidth]{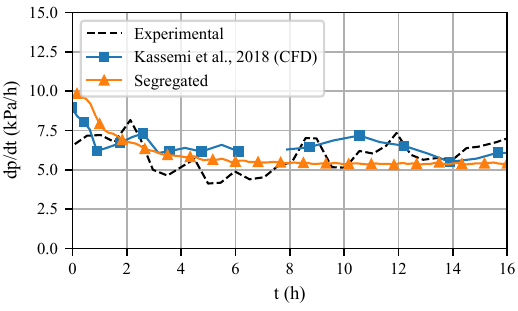}
		\caption{}
		\label{fig:hasan_validation_gradP}
	\end{subfigure}
	\caption{Temporal evolution of the pressure $p(t)$ (a) pressurization rate $dp/dt (t)$ (b) in the large-scale experiments of Hasan, Van Dresar and coworkers \cite{hasan_self-pressurization_1991,van_dresar_self-pressurization_1992}, in the CFD simulations of Kassemi et al. \cite{kassemi_validation_2018}, and the one obtained with our segregated solver. The markers are shown only for plotting purposes.}
	\label{fig:hasan_validation}
\end{figure}

For the large-scale configuration, all simulations were initialized from a preconditioned state with a prescribed vertical temperature gradient in the ullage, consistent with the experimental thermal stratification at $t=0$. Figure~\ref{fig:hasan_validation} compares the pressure evolution and pressurization rates obtained from experiments, from the VOF+KTG simulations of Kassemi et al.~\cite{kassemi_validation_2018}, and from the present segregated solver. The agreement is excellent throughout the simulation, both in terms of pressure levels and pressurization rates, with the segregated solver achieving comparable accuracy without requiring tuning of accommodation coefficients.

\begin{figure}[h]
	\centering
	\begin{subfigure}[a]{0.95\linewidth}
		\includegraphics[width=\linewidth]{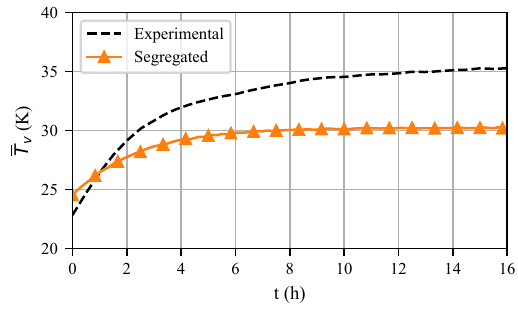}
		\caption{}
		\label{fig:hasan_validation_T}
	\end{subfigure}
	\hfill
	\begin{subfigure}[b]{0.95\linewidth}
		\includegraphics[width=\linewidth]{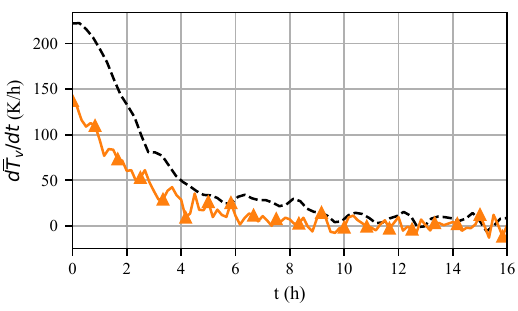}
		\caption{}
		\label{fig:hasan_validation_gradT}
	\end{subfigure}
	\caption{Temporal evolution of the vapor temperature $\overline{T}_v$ (a) and its temporal derivative $d\overline{T}_v /dt$ (b) in the large-scale experiments experiments of Hasan, Van Dresar and coworkers \cite{hasan_self-pressurization_1991,van_dresar_self-pressurization_1992}. The markers are shown only for plotting purposes.}
	\label{fig:hasan_validation_temperatures}
\end{figure}

Figure~\ref{fig:hasan_validation_temperatures} complements this comparison by showing the evolution of the mean vapor temperature and its temporal derivative. Both experiments and simulations exhibit a rapid temperature increase followed by a transition to a quasi-steady regime, in which pressure and temperature rates become nearly constant. This transition occurs at approximately $t\approx10\,\mathrm{h}$ in the experiments and slightly earlier in the simulations. The remaining discrepancy in the mean ullage temperature at the end of the test is on the order of $5\,\mathrm{K}$, suggesting a modest underestimation of the fraction of heat absorbed by the vapor. Similar trends were reported by Stewart and Moder~\cite{Stewart2016}, who simulated the same test case using a VOF+KTG approach with CHT with the tank walls and attributed temperature discrepancies near the tank lid to uncertainties in the representation of the bolted joint and its effective thermal conductance. Kassemi et al.~\cite{kassemi_validation_2018} reported excellent agreement as well, although their comparison was limited to a single measurement location. Overall, the level of agreement obtained here is consistent with the uncertainties inherent in the experimental configurations, including geometric details, thermophysical property variations at cryogenic conditions, and the difficulty of maintaining perfectly controlled boundary conditions over long durations.

In summary, the segregated solver accurately reproduces the pressure and temperature evolution observed in both laboratory-scale and large-scale self-pressurization experiments. While minor discrepancies are observed during the initial transient phase, primarily due to uncertainties in the initial thermal stratification, the agreement in pressurization and temperature rates during the quasi-steady regime is excellent. These results validate the implementation of an energy-based, tuning-free phase-change formulation and demonstrate the capability of the segregated framework to predict self-pressurization dynamics across largely different scales without empirical calibration.

\section{Results}\label{sec:results}

This section examines the physical mechanisms governing self-pressurization in cryogenic tanks using the segregated solver. Section~\ref{sec:results_flow} analyzes the thermodynamic regimes of self-pressurization and quantifies the contribution of individual processes to the pressure rise. Section~\ref{sec:results_buoyancy} focuses on the role of natural buoyancy, with particular emphasis on its influence on the development of thermal stratification in the liquid.

\subsection{The different phases of self-pressurization}\label{sec:results_flow}

This subsection characterizes the relative roles of the physical mechanisms driving pressure build-up in both laboratory-scale and large-scale configurations. To quantify the contribution of each mechanism, the chain rule is applied to the vapor-phase equation of state $p_v(\overline{\rho}_v,\overline{T}_v)$ expressed as a function of mass-averaged quantities ($\overline{\phi} = V^{-1}\int_V \phi\,\mathrm{d}V$),

\begin{equation}
    \frac{dp}{dt} = 
    \underbrace{ 
      \left. \frac{\partial p}{\partial \rho} \right|_T 
      \frac{1}{V_v} \left( \frac{dm_v}{dt} - m_v \frac{1}{V_v} \frac{dV_v}{dt} \right) 
    }_{\text{phase change}}
    +
    \underbrace{ 
      \left. \frac{\partial p}{\partial T} \right|_\rho 
      \frac{d \overline{T}_v}{dt} 
    }_{\text{internal energy}} ,
    \label{eq:pressure_expansion}
\end{equation}
where variations of the ullage volume $dV_v/dt$ are neglected \cite{ludwig_pressure_2013}, consistent with the fixed liquid--vapor interface assumed in the present simulations. The two terms on the right-hand side represent the contributions of phase change (evaporation or condensation) and vapor sensible heating, respectively.

To assess the relative importance of the two terms in \eqref{eq:pressure_expansion}, two pressurization ratios are introduced, namely

\begin{align} \label{eq:eta_evap_heat}
\eta_{\mathrm{pc}} &=
\left.\frac{\partial p}{\partial \rho}\right|_T
\frac{1}{V_v}
\frac{dm_v}{dt}
\left(\frac{dp}{dt}\right)^{-1},
\\[0.5em]
\eta_{\mathrm{ie}} &=
\left.\frac{\partial p}{\partial T}\right|_\rho
\frac{d\overline{T}_v}{dt}
\left(\frac{dp}{dt}\right)^{-1}.
\end{align}

These quantify the fractional contributions of phase change and internal energy variation to the total pressurization rate.

\begin{figure}[!h] 
    \centering
    
    \begin{subfigure}[b]{0.95\linewidth} 
        \centering
        \includegraphics[width=\linewidth]{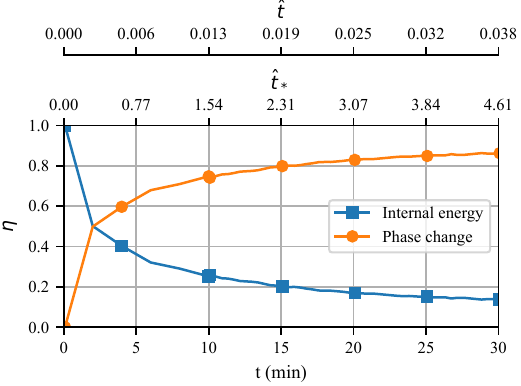}
        \caption{}
        \label{fig:perez_eta_evap_heat}
    \end{subfigure}
    
    \begin{subfigure}[b]{0.95\linewidth}
        \centering
        \includegraphics[width=\linewidth]{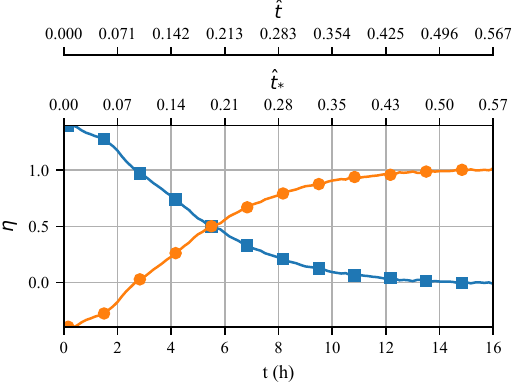}
        \caption{}
        \label{fig:hasan_eta_evap_heat}
    \end{subfigure}
    
    \caption{Temporal evolution of the pressurization ratios $\eta$ computed via Equation~\ref{eq:eta_evap_heat} for the laboratory (a) and large-scale (b) experiments. The time is shown in dimensional (bottom) and dimensionless units (top) using the proposed scaling ($\hat{t}_*$) and the one from Marques et al. \cite{Marques2025} ($\hat{t}$). The markers are shown only for plotting purposes.}
    \label{fig:eta_evap_heat}
\end{figure}

Figure~\ref{fig:eta_evap_heat} shows the temporal evolution of these ratios for both test cases, using dimensional time (bottom axis) and dimensionless time (top axis) based on the proposed ullage-based scaling ($\hat{t}_*$) and the liquid-based scaling of Marques et al.~\cite{Marques2025} ($\hat{t}$). Despite differences in geometry, working fluid, and imposed heat flux, both configurations exhibit two distinct phases of self-pressurization: an initial transient dominated by vapor heating, followed by a quasi-steady regime controlled by evaporation. During the early stage, the internal-energy contribution $\eta_{\mathrm{ie}}$ dominates, indicating that the pressure rise is primarily driven by sensible heating of the ullage. As the system evolves, it transitions to an evaporation-dominated regime in which $\eta_{\mathrm{pc}}$ becomes dominant and both temperature and pressurization rates approach steady values. In the large-scale configuration (Figure~\ref{fig:hasan_eta_evap_heat}), the rapid initial pressure rise is accompanied by a short period of net condensation, indicated by $\eta_{\mathrm{pc}}<0$. This behavior arises because the saturation temperature at the interface increases faster than the bulk liquid temperature as pressure builds up, causing the subcooled liquid to absorb heat from the vapor and induce condensation. At later times, $\eta_{\mathrm{pc}}$ stabilizes at approximately $0.9$ in the laboratory-scale case and close to unity in the large-scale case, reflecting the larger fraction of sensible heat relative to latent heat in the laboratory configuration, consistent with its higher Jakob number.

The duration of the heating-dominated regime differs markedly between the two configurations. In the laboratory-scale case, the system transitions to the evaporation-dominated regime within approximately $5\,\mathrm{min}$, whereas the large-scale experiment requires nearly $6\,\mathrm{h}$. Despite this disparity, the proposed ullage-based scaling captures the transition between regimes consistently, with the switch occurring at $\hat{t}_* \approx 0.2$ in both cases. This observation suggests that the proposed reference time scale provides a meaningful estimate of the duration of the heating-dominated phase and could be used to guide the design of cryogenic storage experiments, for which this initial regime is usually not the primary focus.

\begin{figure}[h!]
	\centering
	\begin{subfigure}[a]{0.95\linewidth}
		\includegraphics[width=\linewidth]{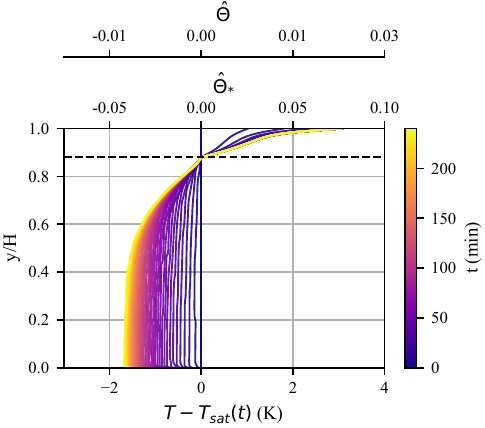}
		\caption{}
		\label{fig:perez_Tprofiles_dless_CFD}
	\end{subfigure}
	\hfill
	\begin{subfigure}[b]{0.95\linewidth}
		\includegraphics[width=\linewidth]{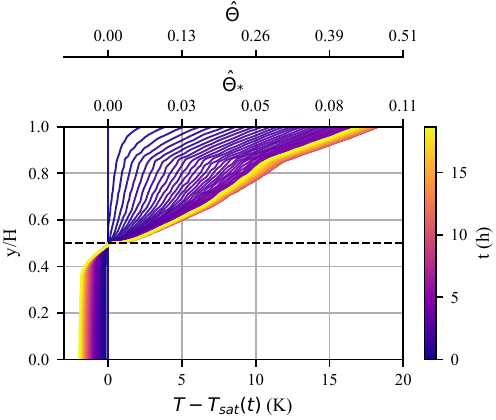}
		\caption{}
		\label{fig:hasan_Tprofiles_dless_CFD}
	\end{subfigure}
	\caption{Temporal evolution of the temperature profiles in the vapor normalized with the saturation temperature ${T} - T_{sat}(t)$ across the $y$-axis normalized with the tank height $H$ in the laboratory (a) and large-scale (b) tests. The top axes show the temperature difference in dimensionless units, using the proposed scaling ($\hat{\Theta}_*$) and the one from Marques et al. \cite{Marques2025} ($\hat{\Theta}$). The interface location is indicated with a dashed black line.}
	\label{fig:Tprofiles_dless}
\end{figure}

Figure~\ref{fig:Tprofiles_dless} shows the evolution of the centerline vapor temperature profiles, expressed relative to the saturation temperature as $T - T_{\mathrm{sat}}(t)$ in order to isolate the thermal evolution of the ullage from the pressure-induced increase in interfacial temperature. As in Figure~\ref{fig:eta_evap_heat}, the upper axes report the corresponding dimensionless temperature using the proposed scaling $\hat{\Theta}_*$ and the liquid-based scaling of Marques et al.~\cite{Marques2025} ($\hat{\Theta}$).

In the vapor phase, the temperature field evolves toward a self-similar profile characterized by an approximately constant vertical gradient. This gradient is established at the wall as a consequence of the imposed heat flux and progressively propagates downward into the ullage. When it reaches the liquid--vapor interface, conduction channels external heat directly into interfacial evaporation, marking the transition to the evaporation-dominated regime. In both configurations, the resulting dimensionless temperature remains within the narrow range $\hat{\Theta}_* \in [0,0.1]$. The development of thermal stratification in the liquid and the mechanisms responsible for it are examined in detail in Section~\ref{sec:results_buoyancy}.

In summary, self-pressurization proceeds through two distinct stages: a rapid, heating-dominated transient followed by a slower, evaporation-dominated quasi-steady regime. These observations support the robustness of the proposed phase-change formulation, which consistently captures the pressurization rate in the evaporation-dominated regime for both configurations, independent of the initial conditions.

The self-similar structure of the vapor temperature profiles observed in the evaporation-dominated regime further enables the derivation of simple analytical relations. Since the mean vapor temperature $\overline{T}_v$ closely follows the saturation temperature, the approximation

\begin{equation} \label{eq:dTdt}
 \frac{d\overline{T}_v}{dt} \approx \frac{dT_{\text{sat}}}{dp} \frac{dp}{dt} \,,
\end{equation} 

can be introduced, where $dT_{\mathrm{sat}}/dp$ is a known thermodynamic property. Substituting this relation into Equation~\ref{eq:pressure_expansion} and rearranging yields the following estimate for the evaporation rate:

\begin{equation}
\frac{d m_v}{dt} \approx 
V_v\left(\left.\frac{\partial p}{\partial \rho}\right|_T\right)^{-1}
\left[1-\left.\frac{\partial p}{\partial T}\right|_\rho \frac{dT_{\mathrm{sat}}}{dp}\right]
\frac{dp}{dt}.
\label{eq:dmdt}
\end{equation}

\begin{figure}[h!]
    \centering
    \begin{subfigure}[b]{0.95\linewidth}
        \centering
        \includegraphics[width=\linewidth]{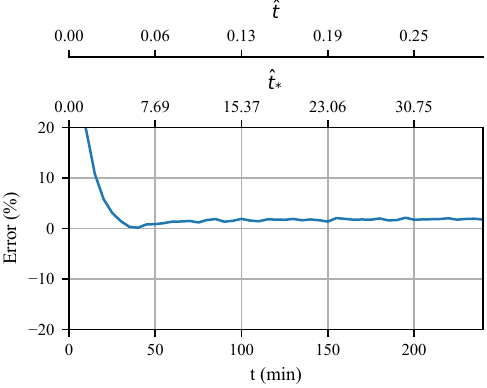} 
        \caption{}
        \label{fig:dmdt_th_error_Perez}
    \end{subfigure}
    
    \begin{subfigure}[b]{0.95\linewidth}
        \centering
        \includegraphics[width=\linewidth]{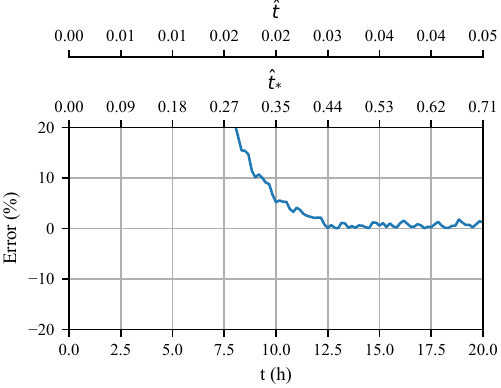}
        \caption{}
        \label{fig:dmdt_th_error_Hasan}
    \end{subfigure}
    \caption{Relative error between the phase change rate $dm/dt$ obtained in the CFD simulations and the one estimated using Equation \ref{eq:dmdt} in the laboratory (a) and large-scale (b) scenarios. The time scale is shown in dimensional (bottom) and dimensionless units (top) using the scaling laws proposed in this work ($\hat{t}_*$) and those proposed in Marques et al. \cite{Marques2025} ($\hat{t}$). The markers are shown only for plotting purposes.}
    \label{fig:dmdt_th_error}
\end{figure}

Figure~\ref{fig:dmdt_th_error} compares the evaporation rate obtained directly from the numerical simulations with the estimate given by Equation~\ref{eq:dmdt}, using the simulated pressurization rate $dp/dt$ as input. Outside the initial heating-dominated phase—where the approximation in Equation~\ref{eq:dTdt} is not expected to hold—the relative error remains below $2\%$ for both test cases. This result indicates that, once the system reaches the evaporation-controlled regime, the evaporation rate can be inferred accurately from pressure measurements alone.

\begin{figure*}[h!]
    \centering
    
    \begin{subfigure}[b]{0.45\linewidth}
        \centering
        \includegraphics[width=\linewidth]{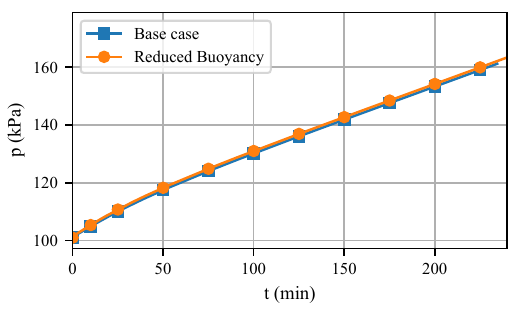}
        \caption{}
        \label{fig:buoyancy_perez_p}
    \end{subfigure}
    \hfill
    \begin{subfigure}[b]{0.45\linewidth}
        \centering
        \includegraphics[width=\linewidth]{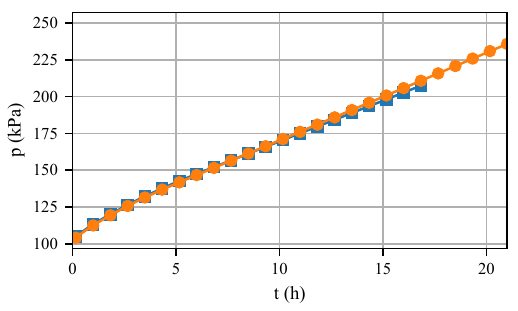}
        \caption{}
        \label{fig:buoyancy_hasan_p}
    \end{subfigure}
    
    \begin{subfigure}[b]{0.45\linewidth}
        \centering
        \includegraphics[width=\linewidth]{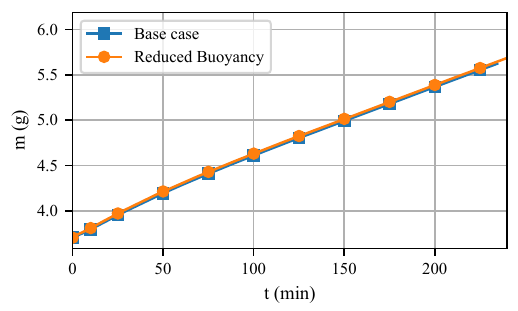}
        \caption{}
        \label{fig:buoyancy_perez_m}
    \end{subfigure}
    \hfill
    \begin{subfigure}[b]{0.45\linewidth}
        \centering
        \includegraphics[width=\linewidth]{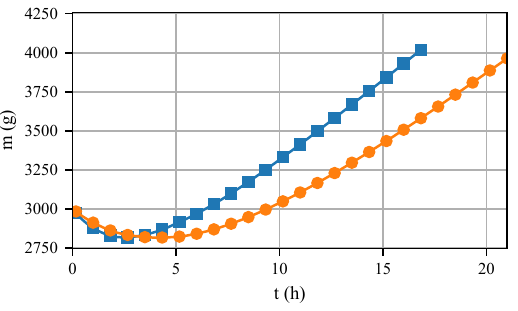}
        \caption{}
        \label{fig:buoyancy_hasan_m}
    \end{subfigure}

    \begin{subfigure}[b]{0.45\linewidth}
        \centering
        \includegraphics[width=\linewidth]{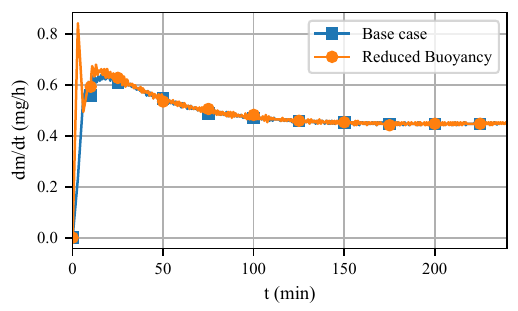}
        \caption{}
        \label{fig:buoyancy_perez_dmdt}
    \end{subfigure}
    \hfill
    \begin{subfigure}[b]{0.45\linewidth}
        \centering
        \includegraphics[width=\linewidth]{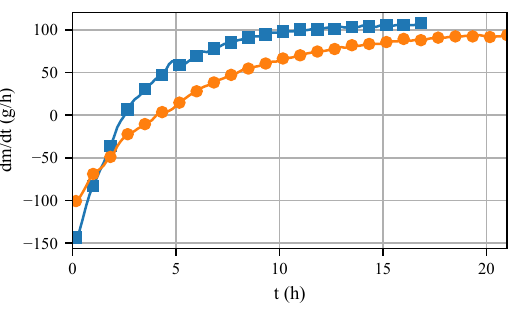}
        \caption{}
        \label{fig:buoyancy_hasan_dmdt}
    \end{subfigure}

    \begin{subfigure}[b]{0.45\linewidth}
        \centering
        \includegraphics[width=\linewidth]{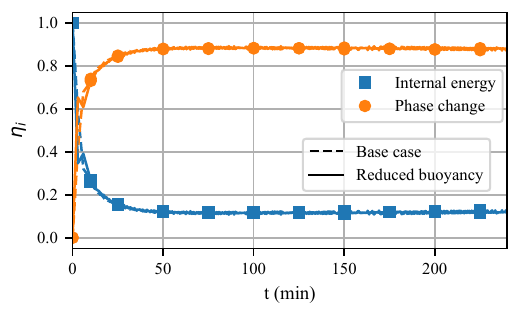}
        \caption{}
        \label{fig:buoyancy_perez_eta}
    \end{subfigure}
    \hfill
    \begin{subfigure}[b]{0.45\linewidth}
        \centering
        \includegraphics[width=\linewidth]{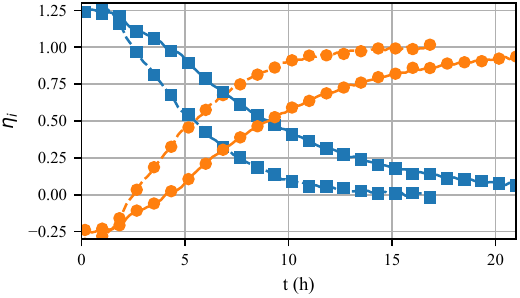}
        \caption{}
        \label{fig:buoyancy_hasan_eta}
    \end{subfigure}
    
    \caption{Comparison between original and reduced buoyancy scenarios for laboratory (left) and large-scale (right) tests. The rows display the temporal evolution of: (a,b) pressure, (c,d) total ullage mass, (e,f) phase-change rate, and (g,h) $\eta_i$, computed via Equation~\ref{eq:eta_evap_heat}. The legends in the left column apply to the corresponding plots on the right, and the markers are shown only for plotting purposes.}
    \label{fig:buoyancy_panel}
\end{figure*}

\subsection{Liquid thermal stratification and buoyancy: the cause or consequence?}
\label{sec:results_buoyancy}

Several studies have investigated the interaction between natural convection and evaporation in cryogenic tanks. Khelifi-Touhami et al.~\cite{KhelifiTouhami2010} characterized buoyancy-driven flow patterns in LNG tanks over a range of Rayleigh numbers and examined their influence on the spatial distribution of interfacial evaporation, while Roh et al.~\cite{Roh2012} analyzed the role of tank geometry. Other works explored passive control strategies, such as wall ribs, to attenuate natural convection in the liquid \cite{Khurana2006,Fu2014}. However, many of these studies assume a constant interfacial temperature, which is not representative of non-venting storage conditions where evaporation induces pressure rise and a concomitant increase in saturation temperature. Under such conditions, the evolving interfacial temperature modifies liquid temperature gradients and can even induce transient condensation, as observed in the large-scale experiments considered here.

To isolate the coupling between natural convection and phase change, additional simulations were performed in which buoyancy terms in the momentum and energy equations were reduced to $10\,\%$ of their nominal values. This approach enables controlled attenuation of convective transport without modifying the imposed heat input, interface geometry, or thermodynamic closure, in contrast to experimental strategies that vary buoyancy indirectly through heat flux \cite{Seo2010,Kang2018} or gravity level \cite{Lv2021}.

Figure~\ref{fig:buoyancy_panel} compares the reference simulations with the reduced-buoyancy cases for both laboratory-scale (left) and large-scale (right) configurations. In the laboratory-scale case, the pressure evolution is essentially insensitive to buoyancy intensity. This behavior is consistent with the rapid transition to the evaporation-dominated regime, in which the system dynamics are governed primarily by the global energy balance rather than by internal convective transport.

In contrast, the large-scale configuration exhibits a noticeable delay in the transient evolution when buoyancy is reduced. Weaker natural convection slows the transport of wall heat into the liquid bulk, thereby extending the heating-dominated phase. Nevertheless, the long-term pressure evolution remains largely unaffected. Figure~\ref{fig:buoyancy_hasan_dmdt} shows that the evaporation rate eventually converges to the same value in both cases, while Figure~\ref{fig:buoyancy_hasan_eta} confirms that the system reaches the same evaporation-dominated state ($\eta_{\mathrm{pc}}\approx1$), albeit on different time scales. These observations indicate that buoyancy primarily controls the duration of the transient phase rather than the asymptotic pressurization rate. As a consequence, passive mitigation strategies aimed at suppressing natural convection (e.g., fins or baffles) are expected to have limited impact on long-term pressure rise, although they may significantly alter transient behavior.

\begin{figure}[h!]
    \centering
    \begin{subfigure}[b]{0.95\linewidth}
        \centering
        \includegraphics[width=\linewidth]{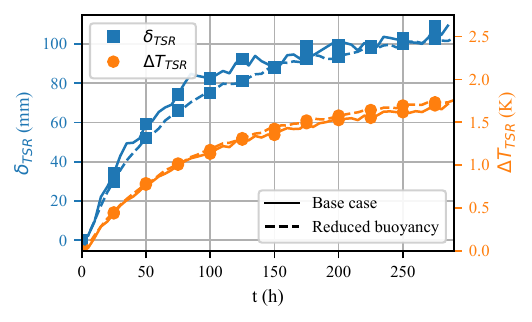}
        \caption{}
        \label{fig:perez_buoyancy_TSR}
    \end{subfigure}
    
    \begin{subfigure}[b]{0.95\linewidth}
        \centering
        \includegraphics[width=\linewidth]{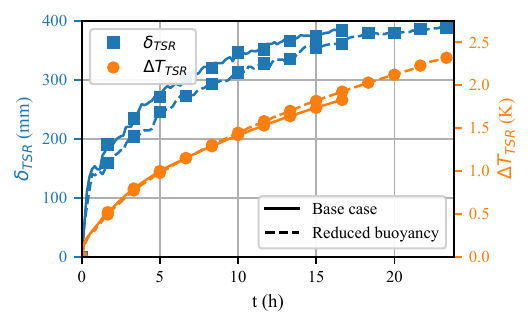}
        \caption{}
        \label{fig:hasan_buoyancy_TSR_sub}
    \end{subfigure}
    
    \caption{Comparison between the original and reduced buoyancy scenarios in terms of the temporal evolution of the width $\delta_{TSR}$ and temperature difference $\Delta T_{TSR}$ of the Thermal Stratified Region (TSR). Results are shown for the (a) lab-scale and (b) large-scale tests. The markers are shown only for plotting purposes.}
    \label{fig:buoyancy_TSR}
\end{figure}

The influence of buoyancy on liquid thermal stratification is examined in Figure~\ref{fig:buoyancy_TSR}, which shows the evolution of the thermal stratified region (TSR), defined as the layer of non-uniform temperature between the liquid--vapor interface at saturation temperature and the nearly isothermal liquid bulk. The thickness of the thermal stratified region (TSR) is defined as
\begin{equation}
\delta_{\mathrm{TSR}}(t) = y_i - y_{\mathrm{TSR}},
\end{equation}
where $y_i$ denotes the interface position. The lower boundary of the TSR, $y_{\mathrm{TSR}}$, is identified as the deepest location in the liquid for which the vertical temperature gradient remains significant,
\begin{equation}
y_{\mathrm{TSR}} =
\max \left\{ y \,\middle|\,
\partial_y T(y,t) > 2.5\,\partial_y T(y_{\mathrm{ref}},t) \right\}.
\label{eq:deltaTSR_definition}
\end{equation}
Here, $y_{\mathrm{ref}}$ is a reference point located in the liquid bulk.

Despite the order-of-magnitude difference in buoyancy forces, the evolution of the thermal stratified region is remarkably similar in both scenarios. The TSR thickness $\delta_{\mathrm{TSR}}$ is only marginally larger in the reference case, while the temperature difference $\Delta T_{\mathrm{TSR}}$ is slightly higher under reduced buoyancy, reflecting weaker mixing in the liquid bulk. This limited sensitivity to convection intensity challenges the traditional view that liquid stratification primarily results from the accumulation of warm fluid transported by buoyancy-driven flows.

Instead, the present results support an alternative interpretation in which TSR development is a thermodynamic consequence of pressure rise. As pressure increases, the saturation temperature at the interface rises accordingly, while the large thermal inertia of the liquid bulk prevents it from responding at the same rate. The stratification therefore develops from the interface toward the bulk, driven by the evolving interfacial boundary condition, rather than from the bottom upward through convective transport. This mechanism is particularly evident in the large-scale configuration, where the initial condensation phase—caused by the lag of the bulk liquid temperature behind the rising saturation temperature—coincides with the rapid establishment of the thermal gradient.

\section{Conclusions and future work}\label{sec:conclusions}

This work investigated the physics of self-pressurization in cryogenic tanks using a segregated numerical framework that couples an incompressible buoyant solver for the liquid phase with a compressible solver for the vapor phase. Phase change at the liquid--vapor interface was modeled through energy-jump conditions enforced as boundary conditions, providing a physics-based alternative to evaporation models that rely on empirically tuned parameters. Conjugate heat transfer between the tank walls and the cryogenic fluid was resolved explicitly, which proved essential given the strong sensitivity of pressurization dynamics to heat partitioning.

The framework was validated against laboratory-scale and large-scale experiments involving liquid nitrogen and liquid hydrogen, respectively. Good agreement was obtained for both pressure and temperature evolution over long durations. Residual discrepancies during the initial transient were attributed primarily to uncertainties in the experimental initial conditions and to differences in thermal inertia between the experimental setups and the idealized numerical models. To the authors’ knowledge, this work presents the first numerical framework capable of predicting self-pressurization across largely different scales without tuning parameters for either interfacial phase change or wall heat transfer, thereby enabling reliable exploration of operating conditions beyond those covered by existing experimental data.

Beyond validation, the simulations provided insight into the mechanisms governing self-pressurization. The process was shown to consist of two distinct regimes: an initial heating-dominated transient, followed by an evaporation-dominated quasi-steady state. During the first regime, the system evolves toward a self-similar temperature profile in the ullage, after which further temperature increase occurs primarily through pressure rise via the saturation relation. Exploiting this behavior, analytical expressions were derived to estimate evaporation rates directly from measured pressurization rates, offering a practical diagnostic tool for situations in which direct measurement of mass transfer is not feasible. An ullage-based scaling introduced in this work successfully collapsed the transition between regimes at $\hat{t}_* \approx 0.2$ and the dimensionless ullage temperature variation within $\hat{\Theta}_* \in [0,0.1]$ across different fluids, tank sizes, geometries, and heat-flux conditions. These results provide a physically grounded basis for designing cryogenic storage experiments and for interpreting pressurization data in long-duration, non-venting scenarios.

The role of natural convection and liquid thermal stratification was examined through additional simulations in which buoyancy forces were systematically reduced. These results showed that buoyancy primarily affects the duration of the heating-dominated transient, while the long-term pressurization and evaporation rates remain largely unchanged. Analysis of the thermal stratified region revealed that stratification growth is governed by the mismatch between the rate of increase of the interfacial saturation temperature—driven by pressure rise—and the slower thermal response of the liquid bulk. This finding contrasts with the conventional view that stratification is primarily built through convective accumulation of warm fluid and highlights the fundamentally thermodynamic origin of stratification in non-venting storage conditions.

Future work will extend the present framework to account for tangential stress continuity at the interface, variable liquid levels, and sloshing dynamics. From a modeling perspective, the segregated formulation also enables independent investigation of turbulence in the liquid and vapor phases, which will be explored in future studies.

\section*{Acknowledgments}

This project was carried out in the framework of the VLAIO grant from the Flemish Government (ref HBC.2023.0897), which also supported David Barreiro. M.A. Mendez is supported by the European Research Council under the European Union’s Horizon Europe research and innovation programme (ERC Starting Grant RE1378 TWIST, Grant Agreement No. 101165479). The authors wish to thank the Centro de Supercomputación de Galicia (CESGA) for providing computational resources, and Prof. Eric May for providing the raw data from the lab-scale experiments.

\section*{Declaration of generative AI and AI-assisted technologies in the writing process}

During the preparation of this work the author(s) used ChatGPT in order to improve the readability of the article. After using this tool/service, the author(s) reviewed and edited the content as needed and take(s) full responsibility for the content of the publication.

\bibliographystyle{elsarticle-num-names}
\bibliography{bibliography}


\end{document}